\documentclass[11pt]{article}
\fontfamily{times}

\usepackage{geometry, amsmath}

\makeatletter
\renewcommand*\env@matrix[1][\arraystretch]{%
  \edef\arraystretch{#1}%
  \hskip -\arraycolsep
  \let\@ifnextchar\new@ifnextchar
  \array{*\c@MaxMatrixCols c}}
\makeatother

\usepackage{bm}
\usepackage{bbm}
\usepackage{epsf}
\usepackage{epsfig}
\usepackage{xcolor}
\usepackage{listings}
\usepackage{amsthm}
\usepackage{mathrsfs}

\usepackage{soul}
\usepackage[style=authoryear,sorting=ynt,natbib=true,bibstyle=numeric,backend=bibtex]{biblatex}
\renewbibmacro{in:}{}
\usepackage{graphicx}
\usepackage{float}
\usepackage{caption}
\usepackage{subcaption}
\usepackage{tikz}
\usetikzlibrary{decorations.markings}
\usetikzlibrary{arrows.meta}
\usepackage{diagbox}
\newtheorem{thm}{Theorem}

\newtheorem{rem}{Remark}
\newtheorem{lemma}{Lemma}

\newcommand{\be}{\begin{equation}}
	\newcommand{\ee}{\end{equation}}

\def\sk1{\vskip 10pt}

\def\uw#1{\underset {^\sim}  {#1}}
\def\wu#1{\overset {_\sim}  #1}

\def\b{{\bold b}}

\def\b0{{\bold 0}}

\def\bold{\bf}

\providecommand{\keywords}[1]{\textbf{\textit{Keywords: }} #1}

\newenvironment{prof}[1][Proof]{\noindent\textit{#1}\quad }
{\hfill $\Box$\vspace{0.7mm}}
 
\usepackage{amssymb} 
\renewcommand{\Bbb}{\mathbb}

\title{Optimal Sequential Procedure for Early Detection \\ of Multiple Side Effects}

\author{Jiayue Wang\thanks{
Email: jwa9@iu.edu } \ and  Ben Boukai\thanks{
Email: bboukai@iu.edu}  \\ Department of Mathematical Sciences, Indiana University Indianapolis\\
Indianapolis, Indiana, 46202 }

\begin{document}
\maketitle

\begin{abstract}

\noindent In this paper, we propose an optimal sequential procedure for the early detection of potential side effects resulting from the administration of some treatment (e.g. a vaccine, say). The results presented here extend previous results obtained in \citet{wang2024early} who study the single side effect case to the case of two (or more) side effects. While the sequential procedure we employ, simultaneously monitors several of the treatment's side effects, the $(\alpha, \beta)$-optimal test we propose does not require any information about the inter-correlation between these potential side effects. However, in all of the subsequent analyses, including the derivations of the exact expressions of the Average Sample Number (ASN), the power function, and the properties of the post-test (or post-detection) estimators, we accounted specifically, for the correlation between the potential side effects. 

In the real-life application (such as post-marketing surveillance), the number of available observations is large enough to justify asymptotic analyses of the sequential procedure (testing and post-detection estimation) properties. Accordingly, we also derive the consistency and asymptotic normality of our post-test estimators; results which enable us to also provide (asymptotic, post-detection) confidence intervals for the probabilities of various side effects. Moreover, to compare two specific side effects, their relative risk plays an important role. We derive the distribution of the estimated relative risk in the asymptotic framework to provide appropriate inference. To illustrate the theoretical results presented, we provide two detailed examples based on the data of side effects on COVID-19 vaccine collected in Nigeria (see \citet{ilori2022acceptability}).

\end{abstract}



\bigskip
\vfill\eject
\setlength{\parindent}{0pt}

\section{Introduction}

In a previous paper, \citet{wang2024early}, we discuss the early direction problem of a single side effect resulting from some treatment applications (e.g., COVID-19 vaccination). We introduced an optimal sequential procedure for such a scenario which matches the fixed sample size of the optimal $(\alpha, \beta)$-UMP test applicable for such a circumstance. In the present paper, we are extending that approach for the early, sequential, detection of two (or more) side effects. Unlike the case of a single side effect, which is captured in a sequential random walk of a single binary process, this generalization, is captured by a sequential random walk over a lattice, of a bivariate binary response whose components are not independent.

In the literature, one may find numerous studies involving, mostly, treatment's efficacy as well as the treatment's safety measure, which may also be viewed as situations involving bivariate 'response' (i.e., 'treatment-response' and say, 'toxicity'), however, in drastically various designs. Some of these studies were carried out as two-stage testing procedures which terminate once certain thresholds on the treatment efficacy {\bf{or}} realized toxicity level have been met. Other were built on a more complex quantitative bivariate response in group sequential settings, see for example, \citet{jennison1993group} who described bivariate normal response in group sequential tests.

A group sequential design for the bivariate binary response was introduced by \citet{conaway1995bivariate} and \citet{conway1995importance} who improved on the calculations via importance sampling. On the other hand, \citet{bryant1995incorporating} constructed two-stage design for bivariate binary response utilizing minimax optimization. Later, \citet{jin2007alternative} modified that design to deal with the trade-off between safety and efficacy. To further reduce the sample-size requirement, \citet{chen2012curtailed} proposed a curtailed two-stage design for two dependent binary responses. \citet{yin2019construction} improved on this stage-wise design with control of the error rates $\alpha$ and $\beta$. We point out that their basic setup is aimed to 'reject' in the null hypothesis that the tested treatment is {\it ineffective or unsafe} against the alternative hypothesis that the treatment is {\it effective and safe}.

In this paper, we introduce a new purely sequential design for the early detection of the multiple potential side effects of a treatment. As a case in point, we highlight the 'treatment' of a vaccination campaign (e.g., COVID-19 vaccination campaign). Nowadays, the development to create new vaccines is being rapidly improving as new technologies are devised and implemented (i.e., mRNA). Further, with the growing need for a quick deployment of the new vaccines to the populous, the concern of several potential side effects of the vaccine becomes a crucial problem.

In modeling such a setup, we first consider the case with two potential side effects. To that end, consider a vaccination campaign for a population of size $N$ (we will expand this later). We assume that the vaccine may potentially cause for two possible side effects which we label as side effect $X$ and side effect $Y$, each realized with some unknown probability (proportion) $\theta_x$ and $\theta_y$, $(0 < \theta_x, \theta_y<1)$, respectively. It is desired to cease the vaccination campaign, if their proportion of the side effects, $\theta_x$ or $\theta_y$ is too large (or unacceptable), namely, if $\theta_x>\theta_x^0$ or $\theta_y>\theta_y^0$, for some 'desired' nominal proportions of these two side effects, $\theta_x^0$ and $\theta_y^0$; otherwise, the vaccination campaign should be continued. This problem can formally be stated as a (sequential) test of the statistical hypotheses:
\begin{equation} \label{1.1}
H_0: \theta_x \le \theta_x^0 \ \text{{\bf{and}}} \ \theta_y \le \theta_y^0  \ \text{against} \ H_1: \theta_x>\theta_x^0 \ \text{{\bf{or}} }\ \theta_y>\theta_y^0.
\end{equation}

For this vaccination campaign, 'patients' sequentially receive the vaccine (a 'treatment' of sorts) and are subsequently classified according to whether or not they have exhibited side effect $X$, side effect $Y$ or both. Clearly, for the sequential testing of the hypotheses \eqref{1.1}, one would terminate the vaccination campaign as soon as it has exhibited too many 'side effects'; otherwise, one would continue uninterruptedly the vaccination process as long as $H_0$ is not rejected.

In Section \ref{s2} below we present the curtailed bivariate sequential test we propose for the above hypotheses \eqref{1.1}. This test is based on a specific stopping rule which hinges on the bivariate binary process. We demonstrate the optimality of the proposed test which achieves the desired probabilities $(\alpha, \beta)$ of the Type~I and Type~II errors, all while meeting the maximal sample-size specification of the respective Uniformly Most Powerful (UMP) test. The derivations and expressions of several important quantities, such as the Average Sample Number (ASN), and the corresponding power function of the test are provided. These derivations account for the fact that the two components of the bivariate binary process, namely $X$ and $Y$ are not independent and hence, the expressions depend on three parameters, $\theta_x$, $\theta_y$, as well as on the correlation between $X$ and $Y$, namely $\rho\equiv \rho_{x,y}$. Since the stopping rule used for the bivariate binary process involves some different termination scenarios, which can be illustrated via a random walk over a lattice, the exact calculations of ASN and the power function are intricate and tedious. However, exploiting some implied relationships between $\theta_x$, $\theta_y$, and $\rho$, we are able to provide some tight bounds for the ASN by simple expressions.

In Section \ref{s4}, for the statistical inference of our optimal bivariate sequential test, we propose the post-test (or post-detection) estimators of $\theta_x$ and $\theta_y$ and analyze their properties. We derive the exact expressions for the expectation of the post-test estimators, again, considering the various possible termination scenarios. Additionally, we further study the properties of our post-test estimators in an asymptotic framework and provide their asymptotic normality. This asymptotic bivariate normality of the post-test estimators is exploited further to simplify the power calculation as well as determination of the ASN. In addition, we discuss in Section \ref{s5} the asymptotic normality of the estimator for the relative risk of these two side effects, and derive the appropriate confidence interval for it. Finally, we close the paper with two detailed examples based on the data of side effects on the COVID-19 vaccine collected from questionnaires in Nigeria (see \citet{ilori2022acceptability}).

\section{The Optimal Bivariate Sequential Test} \label{s2}

Consider a vaccination campaign of a population of $N$ individuals who are being vaccinated sequentially. Each vaccinated individual is being observed for the expression of two possible side effects, labeled here as $X$ and $Y$. Having inspected the first $n$ vaccinated individuals out of $N$, $(1\leq n\leq N)$, the results are summarized in the following $2\times 2$ table.

\begin{table}	
\begin{center}{ 
\caption{Contingency table of $n$ vaccinated people classified by side effect $X$ and side effect $Y$}
\label{t1}
\begin{tabular}{c|c|cc|c}
&  & $Y$  &   &    \  \\ \hline
& 	&  No & Yes  &   \  \\ \hline
$X$  &	No & $n_{00}$ & $n_{01}$  &   \  \\ 
& Yes	& $n_{10}$ & $n_{11}$  &$n_{1+}$ \  \\ \hline
& 	&  &  $n_{+1}$  &  $n$ \  \\ 
\end{tabular}
}
\end{center}
\end{table}
That is, $n_{10}$ of the $n$ vaccinated people have exhibited side effect $X$ only, $n_{01}$ of the vaccinated people have exhibited side effect $Y$ only, whereas $n_{11}$ of the vaccinated people have exhibited both side effects. Clearly, $n_{00}$ of the $n$ vaccinated individuals have exhibited neither of the side effects. For a given $n$, the distribution of these counts is the multinomial distribution,
\begin{equation} \label{2.2}
{\uw{ n}}=(n_{00}, n_{10}, n_{01}, n_{11})^\prime\sim {\mathcal M}N(n, \uw{p}), 
\end{equation}
where $\uw{p}:=(p_{00}, p_{10}, p_{01}, p_{11})^\prime$ is the vector of the corresponding probabilities, $\sum_i\sum_j p_{ij}=1$ along with $\sum_i\sum_j n_{ij}=n$. For each individual, we denote the corresponding classification indicator by ${\uw{z}}^{(k)}= (z_{00,k}, z_{10,k}, z_{01,k}, z_{11,k})^\prime$, where $\sum_i\sum_j z_{ij, k}=1$, for $k=1,2,\dots$. Clearly, 
\begin{equation} \label{2.3}
{\uw{ z}}^{(k)}\sim {\mathcal M}N(1, \uw{ p}), \qquad k=1,2,\dots.
\end{equation}
In view of the sequential nature of the vaccination campaign (and hence of the sampling process), the corresponding sequence of these indicators, ${\uw{ z}}^{(1)}, {\uw{ z}}^{(2)}, \dots$ become available, one–at–a–time or in batches, and the final number of individuals which are being utilized for inference and decisions may depend, in some prescribed fashion, on the data available to the experimenter, namely,  
$$
{\uw{ n}}=\sum_{k=1}^n{\uw{ z}}^{(k)}, \ \ \ \text{and}\ \ \ n_{ij}=\sum_{k=1}^n z_{ij, k}.
$$
Clearly, for each ${\uw{ z}}^{(k)}$, $k=1,\dots,n$, the marginal distributions of the indicator for side effect $X$ (where $X_i=z_{11,i}+z_{10,i}$) and of the indicator for side effect $Y$ (where $Y_j=z_{11,j}+z_{01,j}$) are both Bernoulli random variables, so that,
$$
X \sim {\mathcal B}ern(1, \theta_x), \ \ \ \text{with}\ \  \theta_x:=p_{11}+p_{10},
$$
and
$$
Y\sim {\mathcal B}ern(1, \theta_y), \ \ \ \text{with}\ \  \theta_y:=p_{11}+p_{01}.
$$
Note from the outset that the indicator for the side effect $X$ and the indicator for the side effect $Y$ are not independent random variables (for each $k$). In fact, it can be easily verified (see for example \citet{marshall1985family}), that 
\begin{equation}  \label{2.4}
Cov(X, Y)=p_{11}-\theta_x \theta_y. 
\end{equation}
In the case where the two side effects are statistically independent (rarely), the probabilities $\uw{p}:=(p_{00}, p_{10}, p_{01}, p_{11})^\prime$ are fully specified by the two marginal probabilities $\theta_x$ and $\theta_y$. However, when the two side effects are not independent (more common), the correlation between them is 
$$
\rho=\frac{Cov(X,Y)}{\sqrt{Var(X)Var(Y)}}=\frac{p_{11}-\theta_x \theta_y}{\sqrt{\theta_x(1-\theta_x)\theta_y(1-\theta_y)}}.
$$
Clearly, this correlation $(-1<\rho<1)$, implies some structural restrictions on the parameters and the parameter space. Denoting by $\Omega_x=\theta_x/(1-\theta_x)$ and $\Omega_y=\theta_y/(1-\theta_y)$, the odds of exhibiting side effects $X$ and $Y$, respectively, we have the following three restrictions of $\rho$:
\begin{enumerate}
\item    $p_{11}=\rho \sqrt{\theta_x(1-\theta_x)\theta_y(1-\theta_y)}+\theta_x \theta_y \ge 0 \ \Rightarrow \ \rho  \ge -\sqrt{\Omega_x \Omega_y}$;
\item   $\theta_x-p_{11} \ge 0 \ \Rightarrow \ \rho \le \sqrt{\frac{\Omega_x}{\Omega_y}}$;
\item  $\theta_y-p_{11} \ge 0 \ \Rightarrow \ \rho \le \sqrt{\frac{\Omega_y}{\Omega_x}}$.
\end{enumerate}
Therefore, we have the following condition on the correlation $\rho$.

{\bf{Condition A}}
The values of $\theta_x\in (0,1)$, $\theta_x\in (0,1)$ and the correlation $\rho$, between the two side effects $X$ and $Y$ are such 
$$
\left\{-\sqrt{\Omega_x \Omega_y} \right\}\le \rho \le \min \left\{\sqrt{\frac{\Omega_x}{\Omega_y}}, \sqrt{\frac{\Omega_y}{\Omega_x}}\right\}, 
$$
with $\Omega_x={\theta_x}/({1-\theta_x})$ and $\Omega_y={\theta_y}/({1-\theta_y})$. 

Accordingly, throughout this work, we will restrict attention to the restricted parameter set as defined by
\begin{equation}\label{2.5}
\Theta_R\equiv \left\{ \theta_x\in (0,1), \  \theta_y\in (0,1) \text{ and } \vert \rho \vert <1 \text{ satisfy {\bf{Condition A}}} \right\}.
\end{equation}

Let $S_n^x=X_1+X_2+\cdots+X_n$, denote the number of people exhibited side effect $X$, and $S_n^y=Y_1+Y_2+\cdots+Y_n$, denote the number of people exhibited side effect $Y$. As was stated earlier, large enough values of $S_n^x$ {\bf{or}} $S_n^y$ should lead to the termination of the vaccinated campaign and to some corrective measures (for the patient's treatment). Specifically, in Table \ref{t1}, the values of $S_n^x$ and $S_n^y$ are displayed by $n_{1+}$ and $n_{+1}$, respectively.

Following \citet{wang2024early} (who considered the case of a single side effect), we proceed by obtaining, for given desired probabilities of Type~I error and Type~II error, $(\alpha,\beta)$, the optimal fixed sample size, say $N^*$, and a corresponding critical value $k^*$ for the construction of a UMP test for each of the side effects, separately. For instance, in the case of the side effect $X$, suppose we construct the size $\alpha$ UMP test of $H_0: \theta_x\leq \theta_x^0$ against $H_1: \theta_x> \theta_x^0$, which has a Type~II error probability $\beta$ at some $\theta_x=\theta_x^1>\theta_x^0$. The standard normal approximation of the distribution of $S_n^x$ (see {\bf A.1} in Technical Derivations for details) leads to the calculated values of the optimal $N^*_x$ and a corresponding critical test value $k^*_x$, for the given $(\alpha,\beta,\theta_x^0, \theta_x^1)$. Similarly, for side effect $Y$, one can determine the optimal $N^*_y$ and $k^*_y$ which correspond, to $(\alpha,\beta,\theta_y^0, \theta_y^1)$.

By combining these two separate UMP tests, we consider the construction of an optimal bivariate sequential test of \eqref{1.1} which stops the sampling process as soon as $S^x_n>k^*_x$ {\bf{or}} $S^y_n>k^*_y$. Hence, the sampling (i.e., vaccination) process is to be terminated at a random \emph{stopping time} (see \citet{woodroofe1982nonlinear} for definition), $M=\min \left\{M_x, M_y \right\}$, where 
\begin{equation} \label{2.6}
M_x : =\inf \{n>k^*_x:\ S^x_n>k^*_x\}, \quad M_y :=\inf \{n>k^*_y: \ S^y_n>k^*_y\}.
\end{equation}
The corresponding sequential test of (\ref{1.1}) can be written as:
\begin{equation} \label{2.7}
\operatorname{T_{seq}}: =
\begin{cases}
	\text{if} \ S^x_n=k_x^*+1  \text{ or } S^y_n=k_y^*+1\ & \text{stop and reject $H_0$, $M=n$;} \\
	\text{if }  S^x_n \le k_x^* \text{ and }  S^y_n \le k_y^* \ \  \ & \text{continue the vaccination.} 
\end{cases}
\end{equation}
The following Figure \ref{f1} illustrates the random walk over a lattice of two side effects counting process, which is, the pair $(S_n^x, \, S_n^y)$ jointly defines a random walk over the integer lattice $\{1, \dots n\}^2$.

\begin{figure}[H]
	\caption{The random walk over a lattice}
		\label{f1}
	\centering
	\begin{tikzpicture}[>=Latex,scale=0.5]
		\draw[help lines, color=gray!60](0,0) grid (8,8);
		\fill (0,0) node[left]{$O$} circle (.1);
		\draw[thick,->] (0,0) -- (9,0) node[anchor=north west]{$S_n^{x}$};
		\draw[thick,->] (0,0) -- (0,9) node[anchor=south west]{$S_n^{y}$};
		\draw[semithick,solid,color=brown!60] (0,0) -- (1,1)-- (1,2)-- (2,3)-- (3,3)-- (4,4);
		\fill [color=brown](1,1)  circle (.1);
		\fill [color=brown](1,2)  circle (.1);
		\fill [color=brown](2,3)  circle (.1);
		\fill [color=brown](3,3)  circle (.1);
		\fill [color=brown!60](4,4) circle (.3 );

		\draw[thick,->,color=cyan] (4,4) -- (5,4) node[right,scale=0.8] {$p_{10}$};
		\draw[thick,->,color=cyan] (4,4) -- (4,5) node[above,scale=0.8] {$p_{01}$};
		\draw[thick,->,color=cyan] (4,4) -- (5,5)  node[right,scale=0.8] {$p_{11}$};
		\draw[thick,decoration={markings, mark=at position 0.3 with {\arrow{<}}},
		postaction={decorate},color=blue] (4,3.5) circle(0.5);
		\fill [color=blue](4,3) node[below,scale=0.8]{$p_{00}$} ;
		
		\fill (4,0) node[below,scale=0.8]{$i$} circle (.1);
		\fill (0,4) node[left,scale=0.8]{$j$} circle (.1);
		\fill (5,0) circle (.1);
		\fill (5.5,0) node[below,scale=0.8]{$i+1$}; 
		\fill (0,5) node[left,scale=0.8]{$j+1$} circle (.1);
	\end{tikzpicture}
\end{figure}
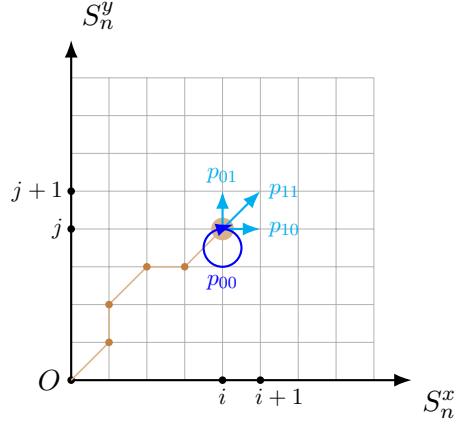

However, since in the most realistic situations, the daily supply of vaccines available to the vaccination center is limited to $N^*:= \min\{N_x^*, N_y^*\}$ units per day (say), the sequential observation (vaccination) process must be terminated once $N^*$ has been reached. Thus, upon termination, the effective random sample size is $M^*:= \min \left\{M,N^*\right\}$. Note that this 'curtailed' sequential test $\operatorname{T^*_{seq}}$ can be written equivalently in terms of the stopping time $M$ as:
\begin{equation} \label{2.8}
\operatorname{T^*_{seq}}: =
\begin{cases}
	\text{if} \ M \le N^* \  \ & \text{stop and reject $H_0$} \\
	\text{if} \ M>N^* \  \ & \text{do not reject $H_0$} 
\end{cases}
\end{equation}

\begin{figure}[H]
	\centering
		\caption{The curtailed sequential test $\operatorname{T^*_{seq}}$}
	\label{f2}
	\begin{tikzpicture}[>=Latex,scale=0.3]
		\draw[help lines, color=gray!60](0,0) grid (20,15);
		\fill (0,0) node[left]{$O$} circle (.1);
		\draw[thick,->] (0,0) -- (22,0) node[anchor=north west]{$S_n^{x}$};
		\draw[thick,->] (0,0) -- (0,17) node[anchor=south west]{$S_n^{y}$};
		\fill (19,0) node[below]{$k^*_x+1$} circle (.2);
		\fill (19,1)  circle (.2);
		\fill (19,2)  circle (.2);
		\fill (19,3)  circle (.2);
		\fill (19,4)  circle (.2);
		\fill (19,5)  circle (.2);
		\fill (19,6)  circle (.2);
		\fill (19,7)  circle (.2);
		\fill (19,8)  circle (.2);
		\fill (19,9)  circle (.2);
		\fill (19,10) node[right]{$(k^*_x+1,k^*_y+1)$} circle (.2);
		\fill [color=brown](20,4) node[right]{-- not reject};
		\fill (20,6) node[right]{-- reject and stop vaccination};
		\fill (0,10) node[left]{$k^*_y+1$} circle (.2);
		\fill (1,10)  circle (.2);
		\fill (2,10)  circle (.2);
		\fill (3,10)  circle (.2);
		\fill (4,10)  circle (.2);
		\fill (5,10)  circle (.2);
		\fill (6,10)  circle (.2);
		\fill (7,10)  circle (.2);
		\fill (8,10)  circle (.2);
		\fill (9,10)  circle (.2);
		\fill (10,10)  circle (.2);
		\fill (11,10)  circle (.2);
		\fill (12,10)  circle (.2);
		\fill (13,10)  circle (.2);
		\fill (14,10)  circle (.2);
		\fill (15,10)  circle (.2);
		\fill (16,10)  circle (.2);
		\fill (17,10)  circle (.2);
		\fill (18,10)  circle (.2);
		\fill (19,10)  circle (.2);
		\draw[semithick,solid] (0,0) -- (1,1)-- (1,2)-- (2,3)-- (3,3)-- (4,4)-- (4,6)--(6,8)--(8,8)--(8,9)--(9,10);
		\fill (1,1)  circle (.1);
		\fill (1,2)  circle (.1);
		\fill (2,3)  circle (.1);
		\fill (3,3)  circle (.1);
		\fill (4,4)  circle (.1);
		\fill (4,5)  circle (.1);
		\fill (4,6)  circle (.1);
		\fill (5,7)  circle (.1);
		\fill (6,8)  circle (.1);
		\fill (7,8)  circle (.1);
		\fill (8,8)  circle (.1);
		\fill (8,9)  circle (.1);
		\draw[semithick,solid,color=brown] (0,0) -- (2,0)-- (4,2)-- (7,2)--(8,3)-- (10,3)-- (10,4)--(12,4);
        \fill [brown] (1,0)  circle (.1);
        \fill [brown](2,0)  circle (.1);
        \fill [brown](3,1)  circle (.1);
        \fill [brown](4,2)  circle (.1);
        \fill [brown](5,2)  circle (.1);
        \fill [brown](6,2)  circle (.1);
        \fill [brown](7,2)  circle (.1);
        \fill [brown](8,3)  circle (.1);
        \fill [brown](9,3)  circle (.1);
        \fill [brown](10,3)  circle (.1);
        \fill [brown](10,4)  circle (.1);
        \fill [brown](11,4)  circle (.1);
        \fill [brown](12,4)  circle (.2);
 
	\end{tikzpicture}
\end{figure}
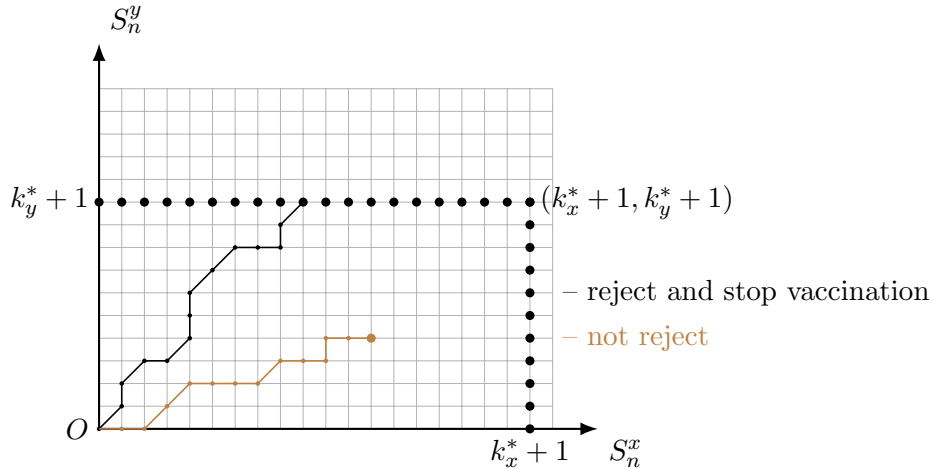

In Figure \ref{f2}, we illustrate the individual path for $S_n^x$ and $S_n^y$, upon rejection (black) and upon non-rejection (brown). 

To study the properties of our bivariate sequential test of \eqref{1.1}, we will consider the power function of the test $\operatorname{T^*_{seq}}$, evaluated at each $\uw{\theta}:=(\theta_x, \theta_y)^\prime\in\Theta_R$,
$$
\Pi_{\operatorname{T^*_{seq}}}(\uw{\theta}):=\operatorname{P}_{\uw{\theta}}(\operatorname{T^*_{seq}} \text{ reject } H_0), 
$$
and the Average Sample Number $ASN^*(\uw{\theta}):=E_{ \uw{\theta}}(M^*)$. 

From \eqref{2.7} and \eqref{2.8}, it follows immediately that, $\forall \ \uw{\theta} \in \Theta_R$,
\begin{equation} \label{2.9}
\Pi_{\operatorname{T^*_{seq}}}(\uw{\theta})=\operatorname{P}_{\uw{\theta}}(M \leq N^* )=1-\operatorname{P}_{\uw{\theta}}(M > N^* )\equiv 1-\operatorname{P}_{\uw{\theta}}\left(S^x_{N^*} \le k_x^* \text{ and }  S^y_{N^*} \le k_y^*\right).
\end{equation}

\begin{thm}\label{thrm1}
Let $\operatorname{T^*_{seq}}$ be the bivariate curtailed sequential test of the hypotheses in \eqref{1.1} as given in \eqref{2.8} above and let $\Pi_{\operatorname{T^*_{seq}}}(\uw{\theta})$ be its power function. For given $(\alpha,\beta)$, we let $\tilde{\alpha}=\alpha/2$ and $\tilde{\beta}=\beta$ be the probabilities of the Type~I and Type~II errors with corresponding $(N_x^*, k_x^*)$ and $(N^*_y, k^*_y)$ for each side effect marginal UMP test with ${\uw{\theta_0}}:=(\theta_x^0,\theta_y^0)$ and $\uw{\theta_1}:=(\theta_x^1,\theta_y^1)$. Then we have that $\operatorname{T^*_{seq}}$ is optimal in the sense that with $N^*=\min\{N_x^*, N_y^*\}$, 
$$
\Pr\left(\text{Type~I error of } \operatorname{T^*_{seq}}\right) \le\Pi_{\operatorname{T^*_{seq}}} (\uw{\theta_0}) \leq \alpha
$$
and 
$$
\Pr\left(\text{Type~II error of } \operatorname{T^*_{seq}}\right) \le 1-\Pi_{\operatorname{T^*_{seq}}} (\uw{\theta_1}) \le \beta.
$$
\end{thm}

\begin{prof} 
In Lemma \ref{lem2} we provide that the power function $\Pi_{\operatorname{T^*_{seq}}}(\uw{\theta})$ is monotonically increasing with respect to $\theta_x$ and $\theta_y$. Hence we have that
$$
\Pr\left(\text{Type~I error of } \operatorname{T^*_{seq}}\right)  \le\Pi_{\operatorname{T^*_{seq}}} (\uw{\theta_0}),
$$
and 
$$
\Pr\left(\text{Type~II error of } \operatorname{T^*_{seq}}\right) \le 1-\Pi_{\operatorname{T^*_{seq}}} (\uw{\theta_1}).
$$
Suppose $N_{x}^* \le N_{y}^*$, we have $N^*=\min\{N_{x}^*, N_{y}^*\}\equiv N_{x}^*$. Note that since $N_{x}^* \le N_{y}^*$, we also have $\operatorname{P}_{\theta_y}(S_{N^*_y}^y> a)=\operatorname{P}_{\theta_y}(\sum_{i=1}^{N^*_y}Y_i> a)\ge\operatorname{P}_{\theta_y}(\sum_{i=1}^{N^*_x}Y_i> a) =\operatorname{P}_{\theta_y}(S_{N^*_x}^y > a) $, $\forall \ a\in {\mathbb{R}}$. Accordingly, by \eqref{2.9}, it follows that   
\begin{align*}
\operatorname{P}_{\uw{\theta_0}} \left(M > N^* \right)=& \operatorname{P}_{\uw{\theta_0}} \left(M_x > N^* \text{ and } M_y > N^* \right) 
=\operatorname{P}_{\uw{\theta_0}} \left(S_{N^*}^x \le k_x^* \text{ and } S_{N^*}^y \le k_y^*  \right) \\
\ge & \operatorname{P}_{\theta_{x}^{0}} \left( S_{N^*}^x \le k_x^* \right)+\operatorname{P}_{\theta_{y}^{0}} \left( S_{N^*}^y \le k_y^* \right) -1 \\
\ge & \left(1-\frac{\alpha}{2}\right)+\left(1-\frac{\alpha}{2}\right) -1 
\ge 1-\alpha. 
\end{align*}
Hence, we conclude that $\Pi_{\operatorname{T^*_{seq}}}(\uw{\theta_0})
=1-\operatorname{P}_{\uw{\theta_0}} \left( M > N^*  \right) \le \alpha$. On the other hand,
\begin{align*}
1-\Pi_{\operatorname{T^*_{seq}}}(\uw{\theta_1})=&1-\operatorname{P}_{\uw{\theta_1}} \left(M \le N^*  \right)  
=\operatorname{P}_{\uw{\theta_1}} \left(S_{N^*}^x \le k_x^* \text{ and } S_{N^*}^y \le k_y^*  \right) \\
\le & \operatorname{P}_{\theta_{x}^{1}} \left(S_{N^*}^x \le k_x^* \right) 
\le  \beta.
\end{align*}
The proof is similar in the case of $N_x^* \ge N_y^*$. This completes the proof of the Theorem. 
\end{prof}

\begin{figure}[H]
\centering
\includegraphics[scale=0.28]{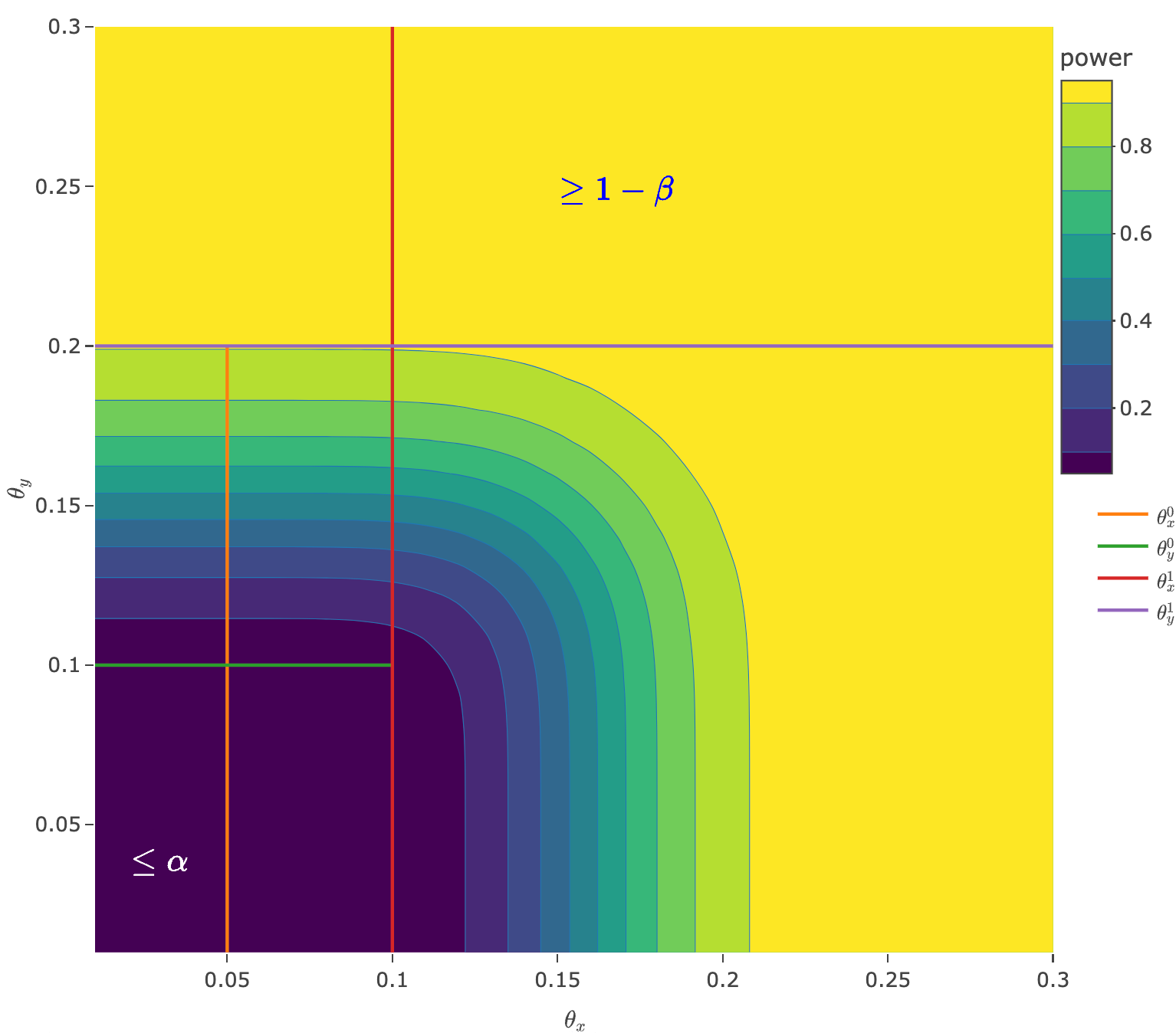}
\caption{The contour plot of the power function with respect to $\theta_x, \ \theta_y \in (0,0.3)$ and $\rho=0.1$ for fixed $\alpha=0.05$, $\beta=0.1$, $\theta_x^0=0.05$, $\theta_x^1=0.1$ and $\theta_y^0=0.1$, $\theta_y^1=0.2$ with $N^*=121$, $k^*_x=19$ and $k^*_y=18$. The corresponding $\Pi_{\operatorname{T^*_{seq}}}(\uw{\theta_0})=0.0321$ and $\Pi_{\operatorname{T^*_{seq}}}(\uw{\theta_1})=0.9065$.}
\label{f3}
\end{figure}

In Figure \ref{f3} below we illustrate the results of Theorem \ref{thrm1} concerning the power function $\Pi_{\operatorname{T^*_{seq}}}(\uw{\theta})$ with respect to the value of $\uw{\theta} \in \Theta_R$. 

\section{On the ASN}\label{s3}

In order to study the efficiency of our proposed optimal bivariate sequential test in (\ref{2.8}), we derive its Average Sample Number (ASN), $ASN^*(\uw{\theta}):=E_{ \uw{\theta}}(M^*)$. Clearly, since $M^*=\min\{M, N^*\}$, we have  
$$
ASN^* (\uw{\theta})= E_{ \uw{\theta}} \left(\min\{M, N^*\} \right) = N^* \operatorname{P}_{\uw{\theta}}\left(M>N^*  \right)+E_{ \uw{\theta}}\left(M \mathbbm{1} \left[M \le N^* \right] \right), 
$$
where $M=\min\{M_x,M_y\}$ and where $\mathbbm{1}[{\mathcal A}]$ is the indicator function of the 'event' ${\mathcal A}$. By Lemma \ref{lem3}, we derive that
\begin{align} \label{3.10}
ASN^* ( \uw{\theta})=& N^* \operatorname{P}_{\uw{\theta}}\left(M>N^* \right)+\sum_{m=1}^{N^*}\operatorname{P}_{\uw{\theta}}\left(M \ge m \right)-N^*\operatorname{P}_{\uw{\theta}}\left(M \ge N^*+1 \right) \notag \\
=&\sum_{m=1}^{N^*}\operatorname{P}_{\uw{\theta}} \left(M \ge m \right):= I^*_1. 
\end{align}
By direct calculation for the term, $I^*_1$, we have that with $\wu{k}:=\min \{k^*_x, k^*_y \}$
\begin{equation} \label{3.11}
I^*_1:= N^*-\sum_{m=\wu{k}+1}^{N^*-1}(N^*-m)\cdot \operatorname{P}_{\uw{\theta}}\left(M=m  \right).
\end{equation}
In Remark \ref{rem3}, we provide the exact calculations of the probabilities of $\operatorname{P}_{\uw{\theta}}(M=m)$ for $m=\wu{k}+1, \dots, N^*$, under $\Theta_R$ in \eqref{2.5}, which once obtained, provide the explicit expression for the evaluation of $ASN^* (\uw{\theta})$ in \eqref{3.10}. These calculations in \eqref{3.11} utilize the ('negative' version of) the multinomial distribution of ${\mathcal M}N(n, \uw{ p})$ in \eqref{2.2}. However, since the exact calculations of the ASN in \eqref{3.11} are intricate, we introduce below some tight bounds for its numerical evaluation.  

To simplify the expressions, we denote, following \eqref{3.10}, 
$$
U_1:=E_{\theta_x}(M^*_x)= \sum_{m=1}^{N^*}\operatorname{P}_{\theta_x}\left(M_x \ge m  \right),  \qquad U_2:=E_{\theta_y}(M^*_y)=\sum_{m=1}^{N^*}\operatorname{P}_{\theta_y}\left(M_y \ge m  \right), 
$$
and 
$$
L_1:=\sum_{m=1}^{N^*}\operatorname{P}_{\theta_x}\left(M_x \ge m \right)\operatorname{P}_{\theta_y}\left(M_y \ge m\right).
$$
Note that it always holds that
$$
L_1\leq \min \left\{U_1, U_2\right\}. 
$$
Also, since $\forall$ $m \in \mathbbm{N}$,
\begin{equation} \label{3.12}
\operatorname{P}_{\uw{\theta}} \left(M \ge m \right)=\operatorname{P}_{\uw{\theta}} \left(\min\left\{M_x,M_y\right\} \ge m  \right)\leq \min \left\{\operatorname{P}_{\theta_x} \left(M_x \ge m \right), \operatorname{P}_{\theta_y} \left(M_y \ge m \right)\right\},
\end{equation}
we obtain that 
$$
\sum_{m=1}^{N^*}\operatorname{P}_{\uw{\theta}} \left(M \ge m  \right) \le \min \left\{U_1, U_2\right\}.
$$
Further, if $M_x$ and $M_y$ are {\it{positively associated}} random variables, then $\forall$ $m \in \mathbbm{N}$,
$$
\operatorname{P}_{\uw{\theta}} \left(M \ge m \right)=\operatorname{P}_{\uw{\theta}}\left(M_x\ge m  \text{ and }M_y \ge m  \right)
 \ge \operatorname{P}_{\theta_x} \left(M_x\ge m \right)\operatorname{P}_{\theta_y}\left(M_y\ge m\right), 
$$
and therefore, 
\begin{equation} \label{3.13}
\sum_{m=1}^{N^*}\operatorname{P}_{\uw{\theta}}\left(M \ge m \right)\ge L_1.
\end{equation}
On the other hand, if $M_x$ and $M_y$ are {\it{negatively associated}} random variables, then $\forall$ $m \in \mathbbm{N}$,
$$
\operatorname{P}_{\uw{\theta}} \left(M \ge m \right)=\operatorname{P}_{\uw{\theta}}\left(M_x\ge m  \text{ and }M_y \ge m \right)
 \le \operatorname{P}_{\theta_x} \left(M_x\ge m \right)\operatorname{P}_{\theta_y}\left(M_y\ge m\right), 
$$
we therefore will have
\begin{equation} \label{3.14}
\sum_{m=1}^{N^*}\operatorname{P}_{\uw{\theta}}\left(M \ge m \right)\le L_1.
\end{equation}
Accordingly, by \eqref{3.12} and \eqref{3.13}, we have the following bounds for the $ASN^* ( \uw{\theta})$
$$
L_1 \le ASN^* (\uw{\theta}) \le \min \left\{U_1, U_2\right\}.
$$
whenever $M_x$ and $M_y$ are positively associated random variables (see Lemma \ref{lem4}). Otherwise, by \eqref{3.14}, we have that 
$$
\wu{k}+1 \le ASN^* (\uw{\theta} ) \le L_1.
$$
Clearly, if $M_x$ and $M_y$ are independent (hence uncorrelated, see Lemma \ref{lem4}), we will have 
$$
ASN^* (\uw{\theta}) =L_1.
$$

Now, since $U_1\equiv ASN^*(\theta_x)$ and $U_2\equiv ASN^*(\theta_y)$, we may obtain, by utilizing Theorem $2$ of \citet{wang2024early} (see expression $(20)$ there for details), that 
$$
U_1=N^*\mathcal{I}_{1-\theta_x}\left(N^*-k^*_x,k^*_x+1\right)+\frac{k^*_x+1}{\theta_x}\mathcal{I}_{\theta_x}\left(k^*_x+2,N^*-k^*_x\right),
$$
where $\mathcal{I}_{\cdot}(\cdot,\cdot)$ denotes the regularized incomplete beta function\footnote{For any $\xi>0, a>0$ and $b>0$, $\mathcal{I}_\xi(a,b):=\int_0^\xi f(u \mid a, b) du\equiv \int_0^\xi \frac{u^{a-1}(1-u)^{b-1}}{b(a, b)}du$, with $f(u \mid a,b)$ being the $pdf$ of a ${\mathcal{B}}eta(a, b)$ random variable. For a detailed discussion of this function and its relation to binomial probabilities, see \citet{hartley1951chart}, \citet{rider1962negative}.}, and similarly, 
$$
U_2=N^*\mathcal{I}_{1-\theta_y}\left(N^*-k^*_y,k^*_y+1\right)+\frac{k^*_y+1}{\theta_y}\mathcal{I}_{\theta_y}\left(k^*_y+2,N^*-k^*_y\right).
$$
For the direct calculation of the expression $L_1$, we must consider two possible situations:
\begin{itemize}
	
\item {(i)}\ \   $k^*_x \ge k^*_y$:
$$
L_1=\mu_{N^*,x}-\sum_{i=k^*_y+1}^{k^*_x}\mathcal{I}_{\theta_y}(k^*_y+1,i-k^*_y)-\sum_{i=k^*_x+1}^{N^*-1}\mathcal{I}_{\theta_y}(k^*_y+1,i-k^*_y)\mathcal{I}_{1-\theta_x}(i-k^*_x,k^*_x+1);
$$

\item {(ii)}\ \   $k^*_x \le k^*_y$:
$$
L_1=\mu_{N^*,y}-\sum_{i=k^*_x+1}^{k^*_y}\mathcal{I}_{\theta_x}(k^*_x+1,i-k^*_x)-\sum_{i=k^*_y+1}^{N^*-1}\mathcal{I}_{\theta_x}(k^*_x+1,i-k^*_x)\mathcal{I}_{1-\theta_y}(i-k^*_y,k^*_y+1).
$$
\end{itemize}

\begin{table}[H]
\begin{center}{\small 
\caption{The values of $ASN^* ( \uw{\theta})$ with respect to $\theta_x$ and $\theta_y$ for fixed $\alpha=0.05$, $\beta=0.1$, $\theta_x^0=0.05$, $\theta_x^1=0.1$ and $\theta_y^0=0.1$, $\theta_y^1=0.2$ with $N^*=121$, $k^*_x=19$ and $k^*_y=18$.}
\label{t2}
\begin{tabular}{c|c|c|c|c|c|c|c|c}
\hline
\multicolumn{2}{c}{} \vline & \multicolumn{7}{c} {$ASN^* ( \uw{\theta})$} \\
\cline{3-9}
\multicolumn{2}{c}{} \vline  & $\left(\theta_x^0, \theta_y^0\right)$ & $\left(\theta_x^1, \theta_y^0\right)$ & $\left(\theta_x^0, \theta_y^1\right)$ & $\left(\theta_x^1, \theta_y^1\right)$ & $(\theta_x^0, 0.25)$ & $(0.25,\theta_y^0)$ & $(0.25,0.25)$ \\
\hline
$\rho=0.1$ & upper & 120.6654 & 120.6654 & 93.8602 & 93.8602 & 75.9630 & 79.9251 & 75.9630 \\
\cline{2-9}
& exact & 120.6653 & 120.5080 & 93.8602 & 93.8397 & 75.9630 & 79.9165 & 69.7126 \\
\cline{2-9}
& lower & 120.6653 & 120.5052 & 93.8602 & 93.8282 & 75.9630 & 79.9095 & 69.2791 \\
\hline
$\rho=-0.1$ & upper & NA & 120.5052 & 93.8602 & 93.8282 & 75.9630 & 79.9095 & 69.2791 \\
\cline{2-9}
& exact & NA & 120.5035 & 93.8602 & 93.8140 & 75.9630 & 79.8995 & 68.8663 \\
\hline
\multicolumn{9}{c}{Note: since $\left(\theta_x^0, \theta_y^0 \right) \notin \Theta_R$ when $\rho=-0.1$, 'NA' is presented.} \\
\end{tabular}
}
\end{center}
\end{table} 
In Table \ref{t2} above we illustrate the calculated bounds for the $ASN^*(\uw{\theta})$ in comparison to its exact calculated value.

After deriving the expression of $ASN^* (\uw{\theta})\equiv E_{\uw{\theta}}(M^*)$, we now consider the second moment, $E_{\uw{\theta}}\left({M^*}^2\right)$, in order to derive $Var_{\uw{\theta}}\left(M^*\right)$.

According Lemma \ref{lem5}, we have that
\begin{align} \label{3.15}
E_{\uw{\theta}}\left({M^*}^2\right)=&{N^*}^2 \operatorname{P}_{\uw{\theta}} \left(M>N^*\right)+E_{\uw{\theta}} \left(M^2 \mathbbm{1}\left[M \le N^*\right]\right) \notag \\
=&{N^*}^2 \operatorname{P}_{\uw{\theta}} \left(M>N^* \right)+2\sum_{m=1}^{N^*}\left(m-\frac{1}{2}\right)\operatorname{P}_{\uw{\theta}}\left(M \ge m \right)-{N^*}^2 \operatorname{P}_{\uw{\theta}} \left(M \ge N^*+1 \right) \notag \\
=&2\sum_{m=1}^{N^*}\left(m-\frac{1}{2}\right)\operatorname{P}_{\uw{\theta}} \left(M \ge m  \right) := I^*_2. 
\end{align}
For the exact calculation of the term, $I^*_2$ above, by direct derivation, we have that
\begin{equation} \label{3.16}
I^*_2:= {N^*}^2-\sum_{m=\wu{k}+1}^{N^*-1}\left({N^*}^2-m^2\right) \cdot \operatorname{P}_{\uw{\theta}} \left(M=m \right).
\end{equation}
Therefore, we may obtain the exact value of $Var_{\uw{\theta}} (M^*)$ by combining the value of $E_{\uw{\theta}}(M^*)$ in \eqref{3.10} and the value of $E _{\uw{\theta}}({M^*}^2)$ in \eqref{3.15}, namely, 
$$
Var_{\uw{\theta}}\left(M^*\right)=E_{\uw{\theta}}\left({M^*}^2\right)-\left[E_{\uw{\theta}}\left(M^*\right)\right]^2=I_2^*-\left(I_1^*\right)^2.
$$
Furthermore, to show the relative dispersion of the terminal sample size, $M^*$, we calculate the value of the coefficient of variation (CV) as 
$$
CV_{\uw{\theta}} \left(M^*\right) := \frac{\sqrt{Var_{\uw{\theta}}\left(M^*\right)}}{E_{\uw{\theta}}\left(M^*\right)}.
$$
As it appears from Table \ref{t3}, the proposed optimal bivariate sequential test results are with relatively low CV with respect to the values of $\theta_x$ and $\theta_y$.

\begin{table}[H]
\begin{center}{\small 
\caption{The values of the variance (and coefficient of variation, CV) of $M^*$ with respect to $\theta_x$ and $\theta_y$ for fixed $\alpha=0.05$, $\beta=0.1$, $\rho=0.1$, $\theta_x^0=0.05$, $\theta_x^1=0.1$ and $\theta_y^0=0.1$, $\theta_y^1=0.2$ with $N^*=121$, $k^*_x=19$ and $k^*_y=18$.}
\label{t3}
\begin{tabular}{|c|c|c|c|c|c|}
\hline
\diagbox{$\theta_x$}{Variance \\ (CV)}{$\theta_y$} & 0.02 & $0.1^*$ & $0.2^*$ & 0.25 & 0.4\\
\hline
0.02 & 3.7e-10 & 6.1436 & 295.2041 & 224.175 & 71.2500\\
& (1.6e-07)  & (0.0205) & (0.1831) & (0.1971) & (0.1777) \\
\hline
$0.05^*$ & 0.0002 & 6.1438 & 295.2041 & 224.1750 & 71.2500 \\
& (0.0001) & (0.0205) & (0.1831) & (0.1971) & (0.1777) \\ 
\hline
$0.1^*$ & 2.7969 & 8.7733 & 294.6476 & 224.0743 & 71.2500 \\
& (0.0138) & (0.0246) & (0.1829) & (0.1971) & (0.1777) \\ 
\hline
0.25 & 232.7980 & 232.4461 & 176.0087 & 139.2098 & 68.8701 \\
& (0.1909) & (0.1908) & (0.1738) & (0.1692) & (0.1751)\\ 
\hline
0.4 & 75.0000 & 75.0000 & 74.0820 & 69.6496 & 43.5088 \\ 
& (0.1732) & (0.1732) & (0.1723) & (0.1680) & (0.1496)\\
\hline
\multicolumn{6}{c}{Note: ${}^*$ denotes values under the hypotheses $H_0$ and $H_1$.} \\
\end{tabular} 
}
\end{center}
\end{table}

\section{Estimation}\label{s4}

Once the optimal sequential test of the hypotheses \eqref{1.1} is implemented, one might be interested in the estimation, upon termination, of the unknown parameters $\theta_x$ and $\theta_y$. In this section, we derive the post-test (or post-detection) estimators $\hat \theta_x$ and $\hat \theta_y$ of the two parameters $\theta_x$ and $\theta_y$ and study their properties (in the finite sample sense as well as in a suitable asymptotic framework).

We begin with the post-test estimator of $\theta_x$, as the derivations of the estimator of $\theta_y$ are similar. Clearly, with the curtailed 'stopping time', $M^*:=\min\{N^*, M\}$ of the sequential test, we consider the 'sample proportion' of those who have exhibited the side effect $X$, namely, 
\be \label{4.17}
\hat{\theta}_{x}=\frac{S_{M^*}^x}{M^*}  \equiv \frac{S_{N^*}^x}{N^*} \mathbbm{1}\left[M>N^*\right] +\frac{S_M^x}{M}\mathbbm{1}\left[M \le N^*\right].
\ee
The expectation of $\hat{\theta}_{x}$ is 
\be \label{4.18}
E_{\uw{\theta}}\left(\hat{\theta}_{x}\right)=E_{\uw{\theta}}\left( \frac{S_{N^*} ^x}{N^*}\mathbbm{1}\left[M>{N^*}\right] \right) +E_{\uw{\theta}}\left(\frac{S_M^x}{M}\mathbbm{1}\left[M \le N^*\right] \right).
\ee
The first term in \eqref{4.18} can be directly calculated by utilizing the multinomial distribution of ${\cal MN}(N^*, \uw{ p})$ in \eqref{2.2} as
\be \label{4.19}
E_{\uw{\theta}}\left(\frac{S_{N^*}^x}{N^*} \mathbbm{1}\left[M>N^*\right]\right)=\sum_{i=0}^{k^*_x-z} \sum_{j=0}^{k^*_y-z} \sum_{z=0}^{\wu{k}}\frac{z+i}{N^*} \frac{N^*!}{z!i!j!(N^*-z-i-j)!}p_{11}^z p_{10}^i p_{01}^j p_{00}^{N^*-z-i-j},
\ee
The second term in \eqref{4.18} is obtained by utilizing the ('negative' version of) the multinomial distribution of ${\cal MN}(n, \uw{ p})$ in \eqref{2.2} as 
\begin{multline} \label{4.20}
E_{\uw{\theta}}\left(\frac{S_M^x}{M}\mathbbm{1}\left[M \le N^*\right] \right)\\
=\sum_{m=k^*_x+1}^{N^*}\sum_{i=1}^{\min\{k^*_x+1,k^*_y\}}\sum_{j=0}^{\min\{k^*_y-i,m-k^*_x-1\}} \frac{k^*_x+1}{m} A^{(1)}_{i,j}
+\sum_{m=k^*_x+1}^{N^*}\sum_{i=0}^{\wu{k}}\sum_{j=0}^{\min\{k^*_y-i,m-k^*_x-1\}}\frac{k^*_x+1}{m} A^{(2)}_{i,j}\\
+\sum_{m=k^*_y+1}^{N^*}\sum_{i=1}^{\min\{k^*_x,k^*_y+1\}}\sum_{j=0}^{\min\{k^*_x-i,m-k^*_y-1\}} \frac{i+j}{m}A^{(3)}_{i,j} 
+\sum_{m=k^*_y+1}^{N^*}\sum_{i=0}^{\wu{k}}\sum_{j=0}^{\min\{k^*_x-i,m-k^*_y-1\}}\frac{i+j}{m} A^{(4)}_{i,j} \\
+\sum_{i=1}^{\wu{k}+1}\sum_{m=k^*_x+k^*_y+2-i}^{N^*} \frac{k^*_x+1}{m} A^{(5)}_{i},
\end{multline}
where the detailed expressions of $A^{(1)}_{i,j},\ A^{(2)}_{i,j},\ A^{(3)}_{i,j},\ A^{(4)}_{i,j},\ A^{(5)}_{i}$ are given in Remark \ref{rem3}.

Therefore, combining \eqref{4.19} and \eqref{4.20} together, we obtain the exact value of the expectation of the post-test estimator $\hat{\theta}_x$ in \eqref{4.18}. In Section \ref{s5}, we discuss the performance of our post-test estimators by way of the maximal relative absolute bias and its tendency between the estimated and the actual value.

\section{Asymptotic Properties}\label{s5}

We study the asymptotic behaviors of our optimal bivariate sequential test in a 'local asymptotic' sense, as $\theta_x^1 \to \theta_x^0$ and $\theta_y^1 \to \theta_y^0$. Specifically, as $\theta_x^1\equiv \theta_x^0(1+\delta)$ and $\theta_y^1\equiv \theta_y^0(1+\delta)$ for some $\delta>0$ and small, with $\delta \rightarrow 0$\footnote{Alternatively, one may take $\theta_x^1=\theta_x^0+\delta$ and $\theta_y^1=\theta_y^0+\delta$, in the subsequent derivations which lead to the same asymptotic results described here.}. Under this parameterization, we denote by $N_x^\delta \equiv N^*_x(\tilde{\alpha}, \beta, \theta_x^0, \delta)$ and $k_x^\delta \equiv k_x^*(\tilde{\alpha}, \beta, \theta_x^0, \delta)$, the optimal $N$ and $k$ of the marginal test on the side effect $X$ and and similarly, by $N_y^\delta$ and $k_y^\delta$ for the marginal test on the side effect $Y$. 

By Lemma $1$ of \citet{wang2024early}, which discussed the case of the single side effect, we have as $\delta\to 0$
$$
N_x^\delta \to \infty, \quad
k_x^\delta \to \infty, \quad 
N_y^\delta \to \infty, \quad 
k_y^\delta \to \infty.
$$
Similarly, we denote by $\wu{k}_\delta :=\min \{k_x^\delta, k_y^\delta \}$. Therefore, we immediately have as $\delta\to 0$,
$$
N^*_\delta= \min\left\{N_x^\delta, N_y^\delta\right\} \to \infty \quad\text{and} \quad  M^*_\delta \ge  \wu{k}_\delta +1 \to \infty.
$$
To approximate the power function $\Pi_{\operatorname{T^*_{seq}}}(\uw{\theta})$ in \eqref{2.9}, we note at first that the power function $\Pi_{\operatorname{T^*_{seq}}}(\uw{\theta})$ can be written in an equivalent form as
$$
\Pi_{\operatorname{T^*_{seq}}}(\uw{\theta})\equiv 1-\operatorname{P}_{\uw{\theta}}\left(M_\delta>N_\delta^*\right)\equiv 1- \operatorname{P}_{\uw{\theta}}\left( S_{N^*_\delta}^x \le k_x^\delta \ \text{and} \ \ S_{N^*_\delta}^y \le k_y^\delta \right).
$$
Further, since $N^*_\delta \to \infty$ as $\delta \to 0$, we may utilize the multivariate version of the CLT, along with lemma \ref{lem2}. It is straightforward to verify that as $\delta \to 0$,
\be \label{5.21}
\begin{pmatrix}[1.5]
	S_{N^*_\delta}^x \\
    S_{N^*_\delta}^y
\end{pmatrix}\sim \mathcal{N}_2\left( \bm{\mu_1},\bm{V_1}\right),
\ee
where $\mathcal{N}_2\left( \bm{\mu_1},\bm{V_1}\right)$ denotes that bivariate normal distribution whose mean and variance-covariance matrix are,  
$$
\bm{\mu_1}=\begin{pmatrix}[1.5]
	N^*_\delta\theta_x \\
	N^*_\delta\theta_y
\end{pmatrix},\quad \bm{V_1}=\begin{pmatrix}[1.5]
	N^*_\delta\theta_x(1-\theta_x) & N^*_\delta\rho\sqrt{\theta_x(1-\theta_x)\theta_y(1-\theta_y) } \\
	N^*_\delta\rho\sqrt{\theta_x(1-\theta_x)\theta_y(1-\theta_y) } & N^*_\delta\theta_y(1-\theta_y) 
\end{pmatrix}.
$$
Accordingly, the power function $\Pi_{\operatorname{T^*_{seq}}}(\uw{\theta})$ can be approximated (when incorporated the standard continuity correction), as $\delta \to 0$, by 
\be \label{5.22}
\Pi_{\operatorname{T^*_{seq}}}(\uw{\theta})= 1-\int_{-\infty}^{k_x^\delta+0.5}\int_{-\infty}^{k_y^\delta+0.5} \phi_2\left(u, w\mid \bm{\mu_1}, \bm{V_1} \right) dw du,
\ee
where $\phi_2(u, w \mid \bm{\mu_1}, \bm{V_1})$ denotes the {\it{pdf}} of the bivariate normal distribution $\mathcal{N}_2\left( \bm{\mu_1},\bm{V_1}\right)$, above. 

The next lemma is a restatement of Theorem $3$ (i) and Theorem $4$ of \citet{gut1983limiting}, which is critical for the derivations of the asymptotic approximation to the $pmf$ of the stopping time $M_\delta$, namely of $\operatorname{P}_{\uw{\theta}}(M_\delta =m)$. 

As defined above, let $\left\{\left(X_i, Y_i\right)\right\}_{i=1}^\infty$ be i.i.d. two-dimensional binary random variables, such that $0<E(X_1)=\theta_x< \infty, \ 0<Var(X_1)=\theta_x (1-\theta_x)<\infty$ and $0<E(Y_1)=\theta_y< \infty, \ 0<Var(Y_1)=\theta_y (1-\theta_y)<\infty$. Further, let $S_n^x=\sum_{i=1}^n X_i,\  S_n^y=\sum_{i=1}^n Y_i$ and let $M_\delta \equiv M^\delta_x \equiv \inf \{n>k_x^\delta: S_n^x>k_x^\delta\}$ (which is denoted as $\tau(t)$ in \citet{gut1983limiting}) be the corresponding stopping time. 

\begin{lemma} \label{lem1}
Let $\eta^2:=Var_{\uw{\theta}}(\theta_y X-\theta_x Y)$, since the joint distribution of $(X,Y)$ can be represented as in \eqref{2.3} of Section \ref{s2}, we have by \eqref{2.4} that,  
$$
\eta^2= \theta_y^2 Var_{\theta_x}\left(X\right)+\theta_x^2 Var_{\theta_y}\left(Y\right)-2\theta_x \theta_y Cov_{\uw{\theta}}\left(X,Y\right)= \theta_x \theta_y\left(\theta_x+\theta_y-2p_{11}\right)>0.
$$
Hence, with $M_\delta \equiv M^\delta_x$, and $\delta \to 0$, 
$$
\frac{S^y_{M_\delta}}{k_x^\delta+1} \xrightarrow{P} \frac{\theta_y}{\theta_x},
$$
and the asymptotic distribution of $(S_{M_\delta}^y,M_\delta)^\prime$ is the normal ${\mathcal{N}}_2(\bm{\mu_2}, \bm{V_2})$,  where 
$$
\bm{\mu_2}=\begin{pmatrix}[1.5]
\frac{\theta_y}{\theta_x} \left(k^\delta_x+1\right) \\
\frac{1}{\theta_x} \left(k^\delta_x+1\right)
\end{pmatrix},\quad \bm{V_2}=\begin{pmatrix}[1.5]
\frac{\theta_y \left(\theta_x+\theta_y-2p_{11}\right)\left(k^\delta_x+1\right)}{\theta_x^2} & \frac{\left(\theta_y-p_{11}\right)\left(k^\delta_x+1\right)}{\theta_x^2} \\
\frac{\left(\theta_y-p_{11}\right)\left(k^\delta_x+1\right)}{\theta_x^2} & \frac{\left(1-\theta_x\right)\left(k^\delta_x+1\right)}{\theta_x^2}
\end{pmatrix}.
$$
Similarly, when $M_\delta \equiv M^\delta_y \equiv \inf \{n>k_y^\delta: S_n^y>k_y^\delta\}$, we obtain as $\delta \to 0$,
$$
\frac{S^x_{M_\delta}}{k_y^\delta+1}  \xrightarrow{P}  \frac{\theta_x}{\theta_y},
$$
and $(S_{M_\delta}^x,M_\delta)^\prime$ is asymptotically normal ${\mathcal{N}}_2(\bm{\mu_3}, \bm{V_3})$, where 
$$
\bm{\mu_3}=\begin{pmatrix}[1.5]
\frac{\theta_x}{\theta_y} \left(k^\delta_y+1\right) \\
\frac{1}{\theta_y} \left(k^\delta_y+1\right)
\end{pmatrix},\quad \bm{V_3}=\begin{pmatrix}[1.5]
\frac{\theta_x \left(\theta_x+\theta_y-2p_{11}\right)\left(k^\delta_y+1\right)}{\theta_y^2} & \frac{\left(\theta_x-p_{11}\right)\left(k^\delta_y+1\right)}{\theta_y^2} \\
\frac{\left(\theta_x-p_{11}\right)\left(k^\delta_y+1\right)}{\theta_y^2} & \frac{\left(1-\theta_y\right)\left(k^\delta_y+1\right)}{\theta_y^2}
\end{pmatrix}.
$$   
\end{lemma}

Now, by utilizing Lemma \ref{lem1}, for small $\delta$ (as $\delta \to 0$), we obtain that $\operatorname{P}_{\uw{\theta}}(M_\delta=m)$, the probability of the stopping time at $m$, for $m=\wu{k}_\delta+1 ,\dots,N^*_\delta$, can be approximated from the asymptotic bivariate normal distributions above, as
\begin{align} \label{5.23}
&\operatorname{P}_{\uw{\theta}}\left( M_\delta=m \right) \equiv\operatorname{P}_{\uw{\theta}}\left(\min\{M_x^\delta,M_y^\delta\}=m \right) \notag\\
=&\int_{-\infty}^{k^\delta_y+0.5}\int_{m-0.5}^{m+0.5} \phi_2\left( u, w \mid \bm{\mu_2}, \bm{V_2} \right) dw d u 
+\int_{-\infty}^{k^\delta_x+0.5}\int_{m-0.5}^{m+0.5} \phi_2\left( u, w \mid \bm{\mu_3}, \bm{V_3} \right)d w d u. 
\end{align}
In Technical Derivations {\bf{A.2}} we provide all details leading to the approximation of $\operatorname{P}_{\uw{\theta}}( M_\delta=m )$ in \eqref{5.23} above. Immediately, by \eqref{5.23}, we can approximate the values of $ASN^*(\uw{\theta})$ in \eqref{3.11} and $E_{\uw{\theta}}({M^*}^2)$ in \eqref{3.16} as $\delta \to 0$. In a similar manner, the power function $\Pi_{\operatorname{T^*_{seq}}}(\uw{\theta})$ in \eqref{5.22} can also be approximated, as $\delta\to 0$, as
$$
\Pi_{\operatorname{T^*_{seq}}}(\uw{\theta})=\int_{-\infty}^{k^\delta_y+0.5}\int_{-\infty}^{N^*_\delta+0.5} \phi_2\left( u, w \mid \bm{\mu_2}, \bm{V_2} \right) dw du
+\int_{-\infty}^{k^\delta_x+0.5}\int_{-\infty}^{N^*_\delta+0.5} \phi_2\left( u, w \mid \bm{\mu_3}, \bm{V_3} \right)d w d u.
$$

\begin{rem} \label{rem1}
Note that as $\delta \to 0$, for our proposed test, we have
$$
\operatorname{P}_{\uw{\theta}} \left(M_x^\delta<M_y^\delta\right)=\int_{-\infty}^{k^\delta_y+0.5}\int_{-\infty}^{N^*_\delta+0.5} \phi_2\left( u, w \mid \bm{\mu_2}, \bm{V_2} \right) dw du,
$$
and 
$$
\operatorname{P}_{\uw{\theta}} \left(M_y^\delta<M_x^\delta\right)=\int_{-\infty}^{k^\delta_x+0.5}\int_{-\infty}^{N^*_\delta+0.5} \phi_2\left( u, w \mid \bm{\mu_3}, \bm{V_3} \right)d w d u.
$$
\end{rem}

Accordingly, by \eqref{5.21} and \eqref{5.23} we can simplify the calculation of the expected value of our post-test (post-detection) estimators, $E_{\uw{\theta}} (\hat\theta_x)$. Since we have, as $\delta \to 0$,
\be \label{5.24}
E_{\uw{\theta}}\left(\frac{S_{N^*_\delta}^x}{N^*_\delta} \mathbbm{1}\left[M_\delta > N^*_\delta\right]\right)=\int_{-\infty}^{k^\delta_x+0.5}\int_{-\infty}^{k^\delta_y+0.5} \frac{S_{N^*_\delta}^x}{N^*_\delta} \phi_2\left( u,w \mid \bm{\mu_1}, \bm{V_1} \right) dw du,
\ee
and
\begin{multline}\label{5.25}
E_{\uw{\theta}}\left(\frac{S_{M^*_\delta}^x}{M^*_\delta} \mathbbm{1}\left[M_\delta \le N^*_\delta\right]\right)=\int_{-\infty}^{k^\delta_y+0.5}\int_{-\infty}^{N^*_\delta+0.5} \frac{k^\delta_x+1}{M_\delta} \phi_2\left( u,w \mid \bm{\mu_2}, \bm{V_2}\right) dw du \\
+\int_{-\infty}^{k^\delta_x+0.5}\int_{-\infty}^{N^*_\delta+0.5} \frac{S_{M_\delta}^x}{M_\delta} \phi_2\left( u,w \mid \bm{\mu_3}, \bm{V_3} \right) dw du,
\end{multline}
we have a simplified expression of $E_{\uw{\theta}}(\hat{\theta}_x)$ by combining \eqref{5.24} and \eqref{5.25} as $\delta \to 0$. And also for $E_{\uw{\theta}}(\hat{\theta}_y)$ as $\delta \to 0$,
\be \label{5.26}
E_{\uw{\theta}}\left(\frac{S_{N^*_\delta}^y}{N^*_\delta} \mathbbm{1}\left[M_\delta > N^*_\delta\right]\right)=\int_{-\infty}^{k^\delta_x+0.5}\int_{-\infty}^{k^\delta_y+0.5} \frac{S_{N^*_\delta}^y}{N^*_\delta} \phi_2\left( u,w \mid \bm{\mu_1}, \bm{V_1}\right) dw du,
\ee
and
\begin{multline}\label{5.27}
E_{\uw{\theta}}\left(\frac{S_{M^*_\delta}^y}{M^*_\delta} \mathbbm{1}\left[M_\delta \le N^*_\delta\right]\right)=\int_{-\infty}^{k^\delta_x+0.5}\int_{-\infty}^{N^*_\delta+0.5} \frac{k^\delta_y+1}{M_\delta} \phi_2\left( u,w \mid \bm{\mu_2}, \bm{V_2}\right) dw du \\
+\int_{-\infty}^{k^\delta_y+0.5}\int_{-\infty}^{N^*_\delta+0.5} \frac{S_{M_\delta}^y}{M_\delta} \phi_2\left( u,w\mid \bm{\mu_3}, \bm{V_3}\right) dw du,
\end{multline}
we have the easier expression of $E_{\uw{\theta}}(\hat{\theta}_y)$ by combining \eqref{5.26} and \eqref{5.27} as $\delta \to 0$.

Reflective of the effects of the 'stopping time' $M^*_\delta$ on the parameter estimates $\hat \theta_x$ (and $\hat \theta_y$), we do not expect these estimators to generally be unbiased. However, to study the question of the possible bias of the post-test (post-detection) estimator (e.g. $\hat{\theta}_x$), we may consider the relative absolute bias between the estimate and the actual value should be 
$$
\frac{\lvert E_{\uw{\theta}}\left(\hat{\theta}_x \right)-\theta_x \rvert}{\theta_x} \cdot 100 \%,
$$
where $\lvert \cdot \rvert$ denotes the absolute value. Additionally, we denote by, 
$\gamma={\theta_x}/{\theta_y}$, the relative risk of the two side effects, $X$ and $Y$, and by $\gamma_0=\theta_x^0/\theta_y^0$, its value under the null hypothesis. 

\begin{figure}[H]
\centering
\begin{subfigure}[b]{0.31\textwidth}
\centering
\includegraphics[width=\textwidth]{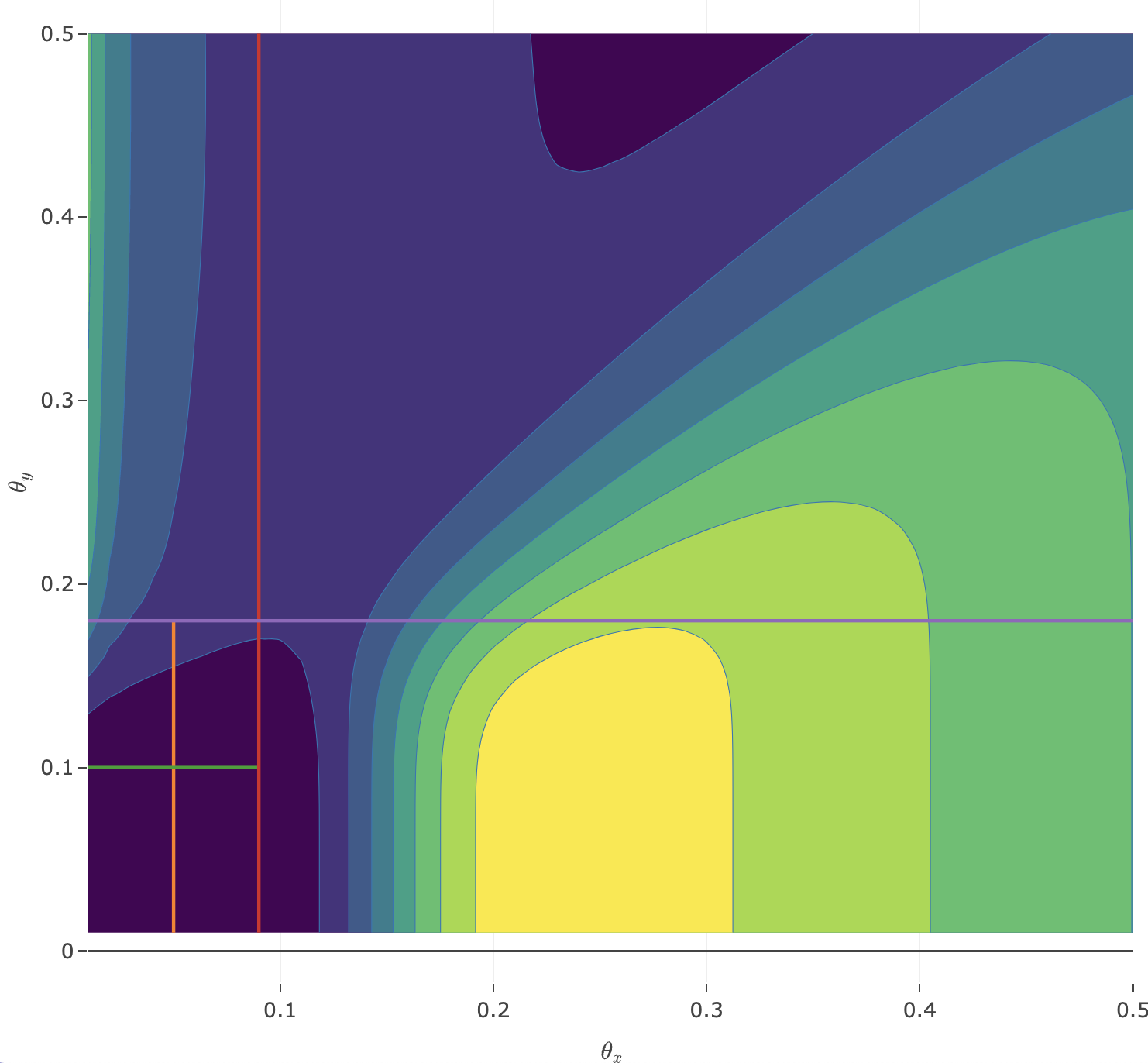}
\caption{$\gamma_0=0.5$}
\end{subfigure}
\hfill
\begin{subfigure}[b]{0.31\textwidth}
\centering
\includegraphics[width=\textwidth]{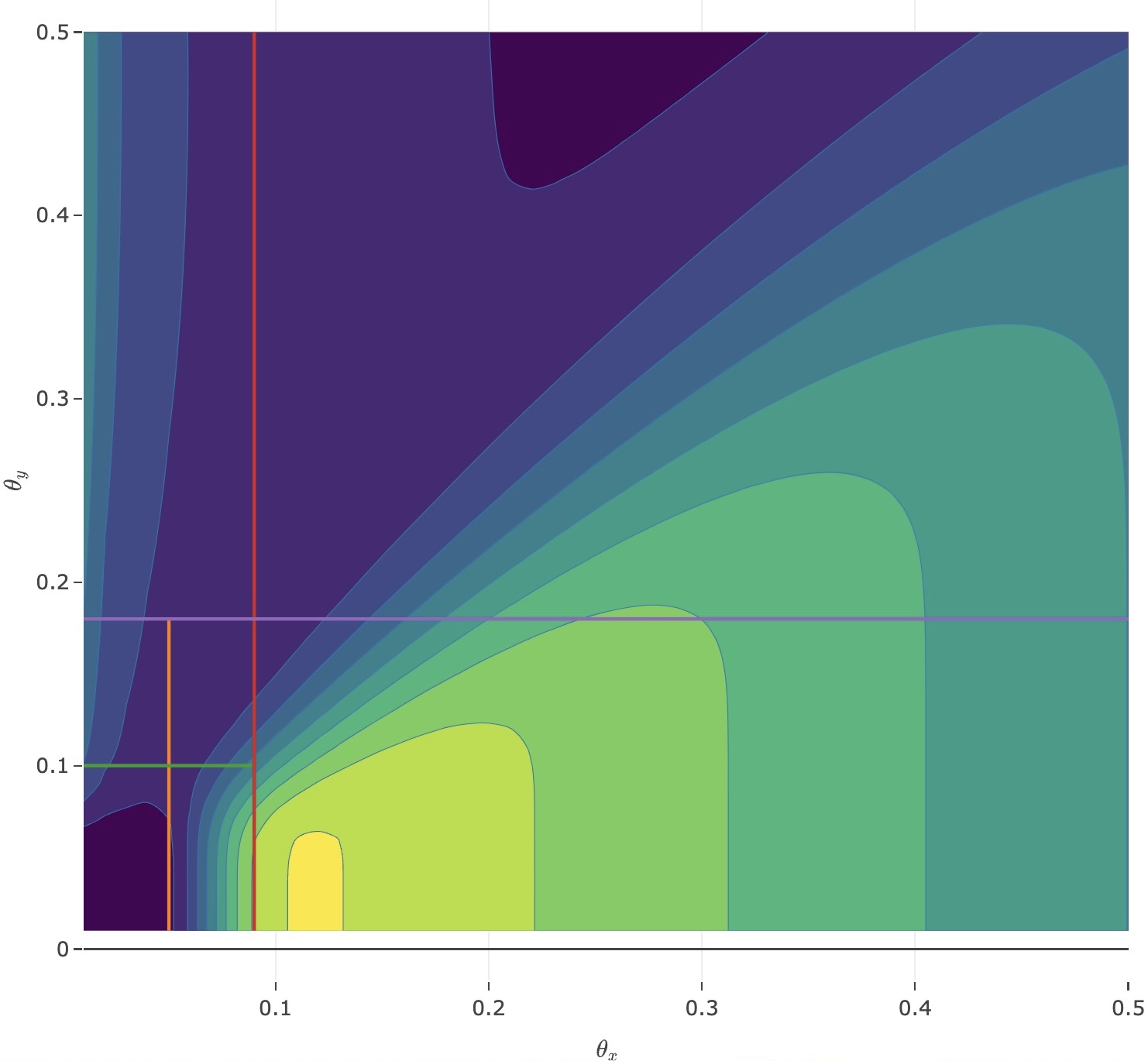}
\caption{$\gamma_0=1$}
\end{subfigure}
\begin{subfigure}[b]{0.365\textwidth}
\centering
\includegraphics[width=\textwidth]{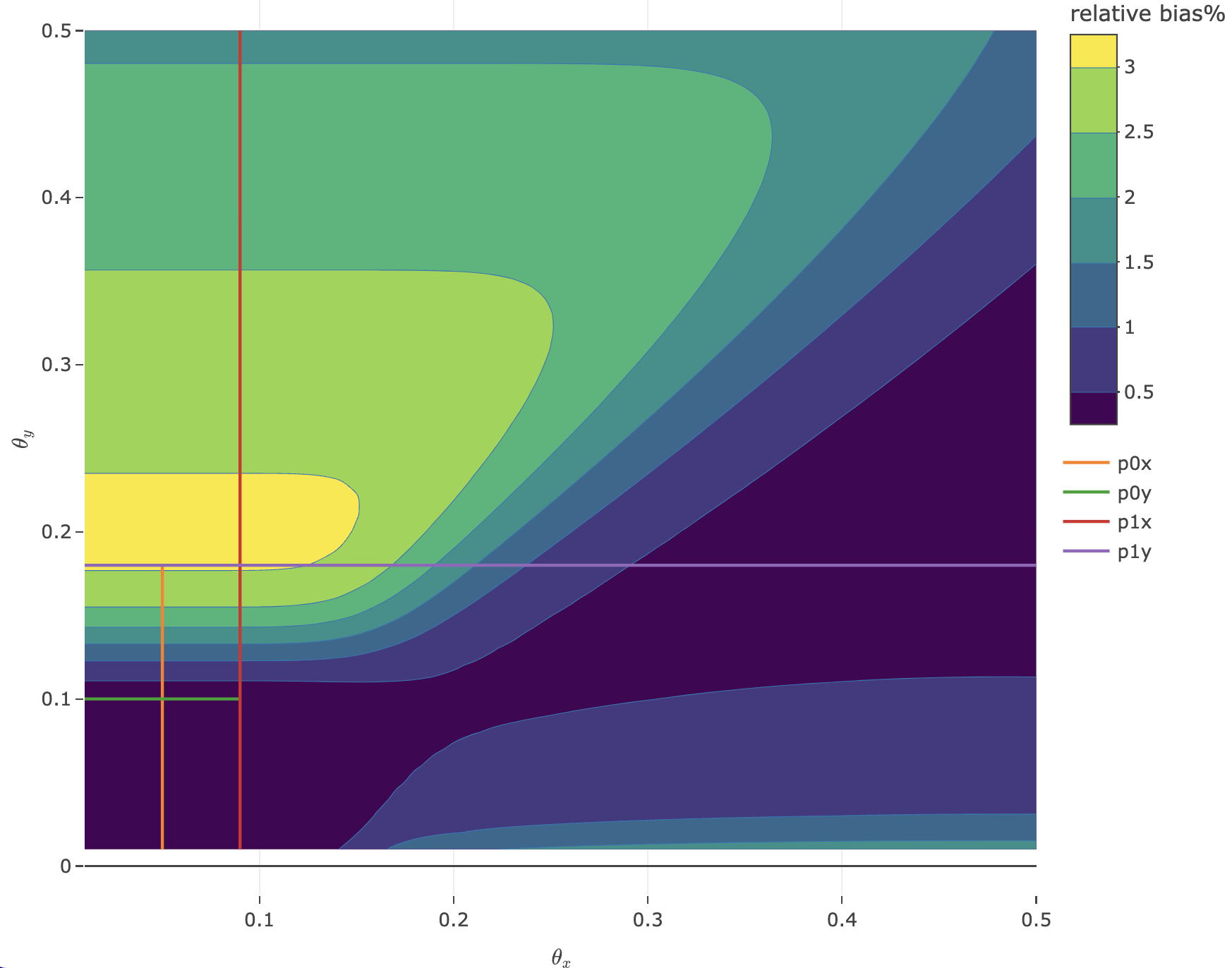}
\caption{$\gamma_0=2$}
\end{subfigure}
\caption{The plots of the tendency of the maximal relative absolute bias among various values of $\theta_x$.}
\label{f4}
\end{figure}

In the following Table \ref{t4} and Figure \ref{f4}, we present the maximal relative absolute bias and its tendency between the estimated and the actual value, which also illustrates the results of Theorem \ref{thrm2} below which indicates the magnitude of the bias tends to $0$. In this illustration, we separate three cases of the relative risk, $\gamma_0=0.5,1,2$. From Figure \ref{f4}, we can conclude that the relative risk does influence the tendency of the maximal relative absolute bias. Since the computing time of the exact calculation based on \eqref{4.18}-\eqref{4.20} substantially increases as the sample size increases, we use the exact calculation only when $\delta > 0.5$ and otherwise we use the Monte Carlo approximation to evaluate \eqref{5.24} and \eqref{5.25}. For instance, from Table \ref{t4}, we can see that when $\delta=0.6$ and $\gamma_0=0.5$, the maximal relative absolute bias is only $1.8528 \%$, which is, in practical terms, very small.

\begin{table}[H]
\begin{center}{\small 
\caption{The percent $(0-100)$ of the maximal relative absolute bias of $E_{\uw{\theta}}(\hat{\theta}_x)$ from exact calculation and Monte Carlo approximation. We assume that (i) $\theta_x^0=0.05$, $\theta_y^0=0.1$;  (ii) $\theta_x^0=\theta_y^0=0.05$;  (iii) $\theta_x^0=0.1$, $\theta_y^0=0.05$. In all cases, $\alpha_x=\alpha_y=0.025$, $\beta_x=\beta_y=0.1$ and $\rho=0.1$.}
\begin{tabular}{c|c|c|c|c|c|c}
\hline
&\multicolumn{6}{c}{maximal relative absolute bias $(\%)$} \\
\cline{2-7}
scenario & $\delta=1$  & $\delta=0.8$ & $\delta=0.6$ & $\delta=0.4$ &   $\delta=0.2$ & $\delta=0.1$ \\
\hline
$\gamma_0=0.5$ & 3.8618 & 2.8326 & 1.8528 & 4.4194  & 2.5015 & 1.3984 \\
\hline
$\gamma_0=1$& 4.5492 & 3.2657  & 2.0921 & 4.4093 & 2.5868 & 1.4513 \\
\hline
$\gamma_0=2$& 4.1259 &  3.1066 & 2.0177 & 4.3121 & 2.4608 & 1.3853  \\
\hline
\end{tabular}
\label{t4}
}
\end{center}
\end{table}

\begin{thm} \label{thrm2}
Let $\hat{\theta}_x$ and $\hat{\theta}_y$ be the post-test (post-detection) estimators (in \eqref{4.17}) of $\theta_x$ and $\theta_y$ based on the $M^*_\delta=\min \left\{N^*_\delta, M_\delta\right\}$ observations obtained from the optimal $(\alpha, \beta)$ bivariate sequential test with $N_\delta^*$, $k_x^\delta$ and $k_y^\delta$. Then as $\delta\to 0$, we have $\forall \ \uw{\theta} \in \Theta_R$, 
$$
\hat{\theta}_x\xrightarrow{P} \theta_x, \quad\text{and} \quad \hat{\theta}_y \xrightarrow{P} \theta_y.
$$
\end{thm}

\begin{prof}
We start with $\hat{\theta}_x$. For small $\delta$ (as $\delta \to 0$), since we can express $\hat{\theta}_x$ by combining the following three terms together, which is
\be \label{5.28}
\hat{\theta}_x=\frac{S^x_{N^*_\delta}}{N^*_\delta} \mathbbm{1}\left[M_\delta>N^*_\delta\right]+\frac{k^\delta_x+1}{M_\delta} \mathbbm{1}\left[M_\delta\le N^*_\delta \text{ and } M_x^\delta< M_y^\delta\right]+\frac{S^x_{M_\delta}}{M_\delta} \mathbbm{1}\left[M_\delta\le N^*_\delta \text{ and }  M_x^\delta> M_y^\delta\right].
\ee

The first term in \eqref{5.28} expresses the case that we would not reject the null hypothesis, in which case $S^x_{N^*_\delta}$ is a binomial process with $N^*_\delta$ and $\theta_x$. Hence, we have
$$
E_{\uw{\theta}} \left(\frac{S^x_{N^*_\delta}}{N^*_\delta}\right)=\theta_x \text{ and } \lim_{\delta \to 0} Var_{\uw{\theta}} \left(\frac{S^x_{N^*_\delta}}{N^*_\delta}\right)=\lim_{\delta \to 0}\frac{\theta_x \left(1-\theta_x\right)}{N^*_\delta}=0.
$$
which indicates that as $\delta \to 0$, $\hat{\theta}_x\xrightarrow{P} \theta_x$ under the case that we would not reject the null hypothesis.

The second term in \eqref{5.28} expresses the case that we would reject the null hypothesis by stopping the process at the boundary $k_x^\delta+1$ corresponding to the side effect $X$ since $S_{M_\delta}^x > k^\delta_x$ and $S_{M_\delta}^y \le k^\delta_y$. From the proof of Theorem $5$ in \citet{wang2024early}, as $\delta \to 0$,
\be \label{5.29}
\hat{\theta}_x \equiv \frac{k^\delta_x+1}{M_\delta} \xrightarrow{P} \theta_x,
\ee
which indicates that if we reject the null hypothesis upon stopping the process at the boundary $k_x^\delta+1$, $\hat{\theta}_x\xrightarrow{P} \theta_x$ as $\delta \to 0$.

The third term in \eqref{5.28} expresses the case that we reject the null hypothesis by stopping the process at the boundary $k_y^\delta+1$ corresponding to the side effect $Y$ since $S_{M_\delta}^y > k^\delta_y$ and $S_{M_\delta}^x \le k^\delta_x$. Since $S_{M_\delta}^y > k^\delta_y$, by the result obtaining in Lemma \ref{lem1}, we have as $\delta \to 0$,
$$
\frac{S_{M_\delta}^x}{k^\delta_y+1} \xrightarrow{P} \frac{\theta_x}{\theta_y}. 
$$
Again, from the proof of Theorem $5$ in \citet{wang2024early}, we have as $\delta \to 0$,
\be \label{5.30}
\frac{k^\delta_y+1}{M_\delta} \xrightarrow{P} \theta_y.
\ee
Hence, we obtain
$$
\hat{\theta}_x\equiv \frac{S_{M_\delta}^x}{M_\delta} \xrightarrow{P} \theta_x, \quad \text{as } \delta\to 0.
$$
Accordingly, in all these three cases, we have as $\delta \to 0$,
$$
\hat{\theta}_x\xrightarrow{P} \theta_x.
$$
In a similar manner, we prove that as $\delta\to 0$, $\forall \ \uw{\theta} \in \Theta_R$, 
$$
\hat{\theta}_y \xrightarrow{P} \theta_y.
$$
\end{prof}

Once Theorem \ref{thrm2} is established, we now introduce one of our main results about the joint asymptotic normality of the post-test (post-detection) estimator $\uw{\hat{\theta}} \equiv (\hat{\theta}_x, \hat{\theta}_y)^\prime$. This result enables us to construct an approximate $(1-\alpha) \cdot 100 \%$ joint confidence interval of the real $\uw{\theta}$. 

\begin{thm} \label{thrm3}
$\forall$ $\uw{\theta} \in \Theta_R$, as $\delta\rightarrow 0$, we have
\be \label{5.31}
\bm{\mathcal {U}}_\delta:=\sqrt{M^*_\delta} \left(\uw{\hat{\theta}}-\uw{\theta}\right) \xrightarrow{D} \mathcal{N}_2
\left(\bm{0},{\bm{\Sigma}}\right),
\ee
where 
$$
{\bm{\Sigma}}=\begin{pmatrix}[1.5]
	\theta_x(1-\theta_x) & p_{11}- \theta_x\theta_y\\
	p_{11}- \theta_x\theta_y & \theta_y(1-\theta_y)
\end{pmatrix}.
$$
\end{thm}

\begin{prof}
We have for any $\uw{t}=\left(t_1,t_2\right)^\prime \in \Bbb{R}^2$, 
\begin{align*}
\operatorname{P}_{\uw{\theta}} \left(\bm{\mathcal {U}}_\delta \leq \uw{t} \right)&= \operatorname{P}_{\uw{\theta}} \left(\bm{\mathcal {U}}_\delta \leq \uw{t} \mid  M_\delta > N^*_\delta \right)\operatorname{P}_{\uw{\theta}} \left(M_\delta > N^*_\delta \right)\\
&+ \operatorname{P}_{\uw{\theta}} \left(\bm{\mathcal {U}}_\delta \leq \uw{t}\mid  M_\delta\leq N^*_\delta \text{ and }  M_x^\delta< M_y^\delta\right)\operatorname{P}_{\uw{\theta}} \left(M_\delta\leq N^*_\delta \text{ and }  M_x^\delta< M_y^\delta \right)\\
&+\operatorname{P}_{\uw{\theta}} \left(\bm{\mathcal {U}}_\delta \leq \uw{t} \mid  M_\delta\leq N^*_\delta \text{ and }  M_x^\delta> M_y^\delta\right)\operatorname{P}_{\uw{\theta}} \left(M_\delta\leq N^*_\delta \text{ and }  M_x^\delta> M_y^\delta \right) \\
&:=B_1 \cdot P_1+B_2 \cdot P_2+B_3 \cdot P_3. 
\end{align*}
Hence, 
\be \label{5.32}
\lim_{\delta \rightarrow 0}\operatorname{P}_{\uw{\theta}} \left( \bm{{\mathcal U}}_\delta \leq \uw{t} \right)=\lim_{\delta \rightarrow 0}B_1\cdot \lim_{\delta \rightarrow 0}P_1 
+\lim_{\delta \rightarrow 0}B_2\cdot \lim_{\delta \rightarrow 0}P_2 
+\lim_{\delta \rightarrow 0}B_3\cdot \lim_{\delta \rightarrow 0}P_3 .  
\ee
Note that $P_1+P_2+ P_3 \to 1$, as $\delta \to 0$. The term $B_1$ in \eqref{5.32} expresses the case that we would not reject the null hypothesis, which means $M^*_\delta \equiv N^*_\delta$ observations. In which case, we have $\uw{\hat{\theta}}=(\frac{S_{N^*_\delta}^x}{N^*_\delta},\frac{S_{N^*_\delta}^y}{N^*_\delta})^\prime$. By the asymptotic distribution of $(S_{N^*_\delta}^x,S_{N^*_\delta}^y)^\prime$ in \eqref{5.21}, we immediately obtain that
$$
\bm{\mathcal {U}}_\delta \equiv \sqrt{N^*_\delta} \left( \uw{\hat{\theta}}-\uw{\theta} \right) \xrightarrow{D} \mathcal{N}_2 \left(
\bm{0},{\bm{\Sigma}} \right), \quad \text{as} \quad \delta \to 0.
$$
Considering now the second term, $B_2$, in \eqref{5.32}, we notice that we would stop early at the boundary $S_{M_\delta}^x=k^\delta_x+1$ corresponding to the side effect $X$, in which case, the stopping time $M_\delta \equiv M_x^\delta$.

By utilizing the asymptotic bivariate normality of $(S_{M_\delta}^y,M_\delta)^\prime$ in Lemma \ref{lem1}, we may derive the distribution of 
$$
\uw{\hat{\theta}} \equiv g_1 \left(
S_{M_\delta}^y, M_\delta \right) :=
\begin{pmatrix}[1.5]
	\frac{k^\delta_x+1}{M_\delta} \\
	\frac{S_{M_\delta}^y}{M_\delta}
\end{pmatrix} ,
$$
by applying the standard delta method. Accordingly, we compute the corresponding gradient vector of the function $g_1$ and evaluate it at $\bm{\mu_2}$.
Therefore, the asymptotic variance is
$$
\nabla g_1   \bm{V_2} {\nabla g_1}^\prime =\begin{pmatrix}[1.5]
\frac{\theta_x^2 (1-\theta_x)}{k^\delta_x+1} & \frac{\theta_x \left(p_{11}- \theta_x\theta_y\right)}{k^\delta_x+1} \\
\frac{\theta_x \left(p_{11}- \theta_x\theta_y\right)}{k^\delta_x+1} & \frac{\theta_x \theta_y(1-\theta_y)}{k^\delta_x+1}
\end{pmatrix}.
$$
Hence, we obtain that as $\delta \to 0$,
$$
\sqrt{\frac{k^\delta_x+1}{\theta_x}}\left( \uw{\hat{\theta}}-\uw{\theta}\right) \xrightarrow{D} \mathcal{N}_2
\left(\bm{0}, {\bm{\Sigma}} \right), \quad \text { as } \quad {\delta} \rightarrow 0.
$$
By the result stated in \eqref{5.29}, we conclude that in this case (when $M_\delta\leq N^*_\delta$ and $M_x^\delta< M_y^\delta$),
$$
\bm{\mathcal {U}}_\delta \equiv \sqrt{M_\delta} \left(\uw{\hat{\theta}} -\uw{\theta}\right)\xrightarrow{D} \mathcal{N}_2
\left(\bm{0},{\bm{\Sigma}}\right), \quad \text { as } \quad {\delta} \rightarrow 0.
$$
To consider the third term $B_3$ in \eqref{5.32}, we notice that we would stop early at the boundary $S_{M_\delta}^y=k^\delta_y+1$ corresponding to the side effect $Y$, in which case, the stopping time $M_\delta \equiv M_y^\delta$. 

By utilizing the asymptotic bivariate normality of $(S_{M_\delta}^x,M_\delta)^\prime$ in Lemma \ref{lem1}, we may derive the distribution of
$$
\uw{\hat{\theta}} \equiv g_2 \left(
S_{M_\delta}^x, M_\delta \right) :=
\begin{pmatrix}[1.5]
	\frac{S_{M_\delta}^x}{M_\delta} \\
	\frac{k^\delta_y+1}{M_\delta}
\end{pmatrix},
$$
by applying the standard delta method. Accordingly, we compute the corresponding gradient vector of the function $g_2$ and evaluate it at $\bm{\mu_3}$.
Therefore, the asymptotic variance is, 
$$
\nabla g_2   \bm{V_3} {\nabla g_2}^\prime =\begin{pmatrix}[1.5]
\frac{\theta_x \theta_y(1-\theta_x)}{k^\delta_y+1} & \frac{\theta_y \left(p_{11}- \theta_x\theta_y\right)}{k^\delta_y+1} \\
\frac{\theta_y \left(p_{11}- \theta_x\theta_y\right)}{k^\delta_y+1} & \frac{\theta_y^2 (1-\theta_y)}{k^\delta_y+1}
\end{pmatrix}.
$$
Hence, we obtain that as $\delta \to 0$,
$$
\sqrt{\frac{k^\delta_y+1}{\theta_y}}\left( \uw{\hat{\theta}}-\uw{\theta}\right) \xrightarrow{D} \mathcal{N}_2
\left(\bm{0}, {\bm{\Sigma}} \right), \quad \text { as } \quad {\delta} \rightarrow 0.
$$
By the result stated in \eqref{5.30}, we conclude that in this case (when $M_\delta\leq N^*_\delta$ and $M_x^\delta> M_y^\delta$),
$$
\bm{\mathcal {U}}_\delta \equiv \sqrt{M_\delta} \left(\uw{\hat{\theta}} -\uw{\theta}\right)\xrightarrow{D} \mathcal{N}_2
\left(\bm{0},{\bm{\Sigma}}\right), \quad \text { as } \quad {\delta} \rightarrow 0.
$$
Hence, upon combining the above together (and accounting of $P_1+P_2+P_3 \to 1$ as $\delta \to 0$), we obtain
\begin{align*} 
\lim_{\delta \rightarrow 0}\operatorname{P}_{\uw{\theta}} \left( {\cal U}_\delta \leq \uw{t} \right) 
&=\Phi_2 \left(\uw{t} \mid \bm{0},{\bm{\Sigma}}\right)\cdot \lim_{\delta \rightarrow 0}P_1 +\Phi_2 \left(\uw{t} \mid \bm{0},{\bm{\Sigma}}\right)\cdot\lim_{\delta \rightarrow 0}P_2
+\Phi_2 \left(\uw{t} \mid \bm{0},{\bm{\Sigma}}\right)\cdot\lim_{\delta \rightarrow 0}P_3 \\
&=\Phi_2 \left(\uw{t} \mid \bm{0},{\bm{\Sigma}}\right),
\end{align*}
where $\Phi_2(\uw{t} \mid \bm{0},{\bm{\Sigma}})$ denotes the joint $cdf$ of the bivariate normal distribution $\mathcal{N}_2(\bm{0},{\bm{\Sigma}})$. Accordingly, we have that
$$
\sqrt{M^*_\delta} \left(\uw{\hat{\theta}} -\uw{\theta}\right) \xrightarrow{D} \mathcal{N}_2
\left(\bm{0},{\bm{\Sigma}}\right) \quad \text { as } \quad {\delta} \rightarrow 0,
$$
as stated in \eqref{5.31}.
\end{prof}

We close this section with a brief discussion of the (asymptotic) properties of the estimated relative risk between the two side effects, $X$, and $Y$, namely of $\gamma:=\theta_x/\theta_y$. Indeed, the following result is an immediate consequence to the results stated in Theorem \ref{thrm3} and it deals with the asymptotic distribution of $\hat \gamma := {\hat \theta_x}/{\hat \theta_y}$.   

\begin{thm} \label{thrm4}
Let $\hat \gamma$ be as above. Then $\forall \ \uw{\theta} \in \Theta_R$, as $\delta\rightarrow 0$,  we have
$$
\sqrt{M^*_\delta}\left( \hat{\gamma}-\gamma\right) \xrightarrow{D} \mathcal{N}
\left(0, \gamma \left(\frac{\gamma+1}{\theta_y}-\frac{2p_{11}}{\theta_y^2}\right)\right).
$$
\end{thm}

\begin{prof}[Outline of the Proof]
Since $\hat{\gamma} \equiv h(\hat{\theta}_x,\hat{\theta}_y)$, by applying the standard delta method to the result in Theorem \ref{thrm3}, we compute the corresponding gradient vector of the function $h$ and evaluate it at $\uw{\theta}$. Therefore, the asymptotic variance is
$$
\nabla h {\bm{\Sigma}}{\nabla h}^\prime=\frac{\theta_x(\theta_x+\theta_y-2p_{11})}{\theta_y^3} \equiv \gamma \left(\frac{\gamma+1}{\theta_y}-\frac{2p_{11}}{\theta_y^2}\right).
$$
Hence, the stated result follows. 

Specifically,  when we stop at the boundary at $k^\delta_x+1$ corresponding to the side effect $X$, by \eqref{5.29}, we have as $\delta \to 0$,
$$
\sqrt{k^\delta_x+1} \left(\hat{\gamma}-\gamma\right) \xrightarrow{D} \mathcal{N}\left(0, \gamma^2 \left(\gamma+1-\frac{2p_{11}}{\theta_y}\right)\right);
$$
when we stop at the boundary at $k^\delta_y+1$ corresponding to the side effect $Y$, by \eqref{5.30}, we have as $\delta \to 0$,
$$
\sqrt{k^\delta_y+1} \left(\hat{\gamma}-\gamma\right) \xrightarrow{D} \mathcal{N}\left(0, \gamma \left(\gamma+1-\frac{2p_{11}}{\theta_y}\right)\right).
$$
\end{prof}

\begin{rem} \label{rem2}
Similarly, denote $\nu:=1/\gamma$. We may obtain the asymptotic normality of $\hat \nu:=1/\hat{\gamma}$. That is, $\forall \ \uw{\theta} \in \Theta_R$, as $\delta \to 0$,
$$
\sqrt{M^*_\delta}\left( {\hat{\nu}}-{\nu}\right) \equiv \sqrt{M^*_\delta}\left( \frac{1}{\hat{\gamma}}-\frac{1}{\gamma}\right) \xrightarrow{D} \mathcal{N}
\left(0,  \nu \left(\frac{\nu+1}{\theta_x}-\frac{2p_{11}}{\theta_x^2}\right)\right).
$$
\end{rem}

\section{Analysis of Some COVID-19 Side Effects Data}\label{s6}

\citet{ilori2022acceptability} provided the data on the side effects to COVID-19 vaccine which were recorded among some health care workers in Nigeria. Their study accounted for $117$ participants who received the COVID-19 vaccine. These vaccinated participants reported on several side effects, if any. In the following Table \ref{t5}, we provide the counts of some of the reported side effects, as fever, muscle pain, dizziness, headache, etc.

\begin{table}[H]
 \centering
 \subfloat[example $1$]{
 	\begin{tabular}{c|c|cc|c}
 		&  &fever &   &    \  \\ \hline
 		& 	&  No & Yes  &   \  \\ \hline
 		muscle pain  &	No & 63 & 11  &   \  \\ 
 		& Yes	& 18 & 25  & 43 \  \\ \hline
 		& 	&  &  36  &  117 \  \\ 
 \end{tabular}}
\hspace{1cm}
\subfloat[example $2$]{
	\begin{tabular}{c|c|cc|c}
		&  &dizziness &   &    \  \\ \hline
		& 	&  No & Yes  &   \  \\ \hline
		headache  &	No & 78 & 5  &   \  \\ 
		& Yes	& 26 & 8  & 34 \  \\ \hline
		& 	&  &  13  &  117 \  \\ 
\end{tabular}}
\caption{$2 \times 2$ table of example $1$ and example $2$}
\label{t5}
\end{table}

\begin{figure}[H]
\centering
\begin{subfigure}[b]{0.48\textwidth}
\centering
\includegraphics[width=\textwidth]{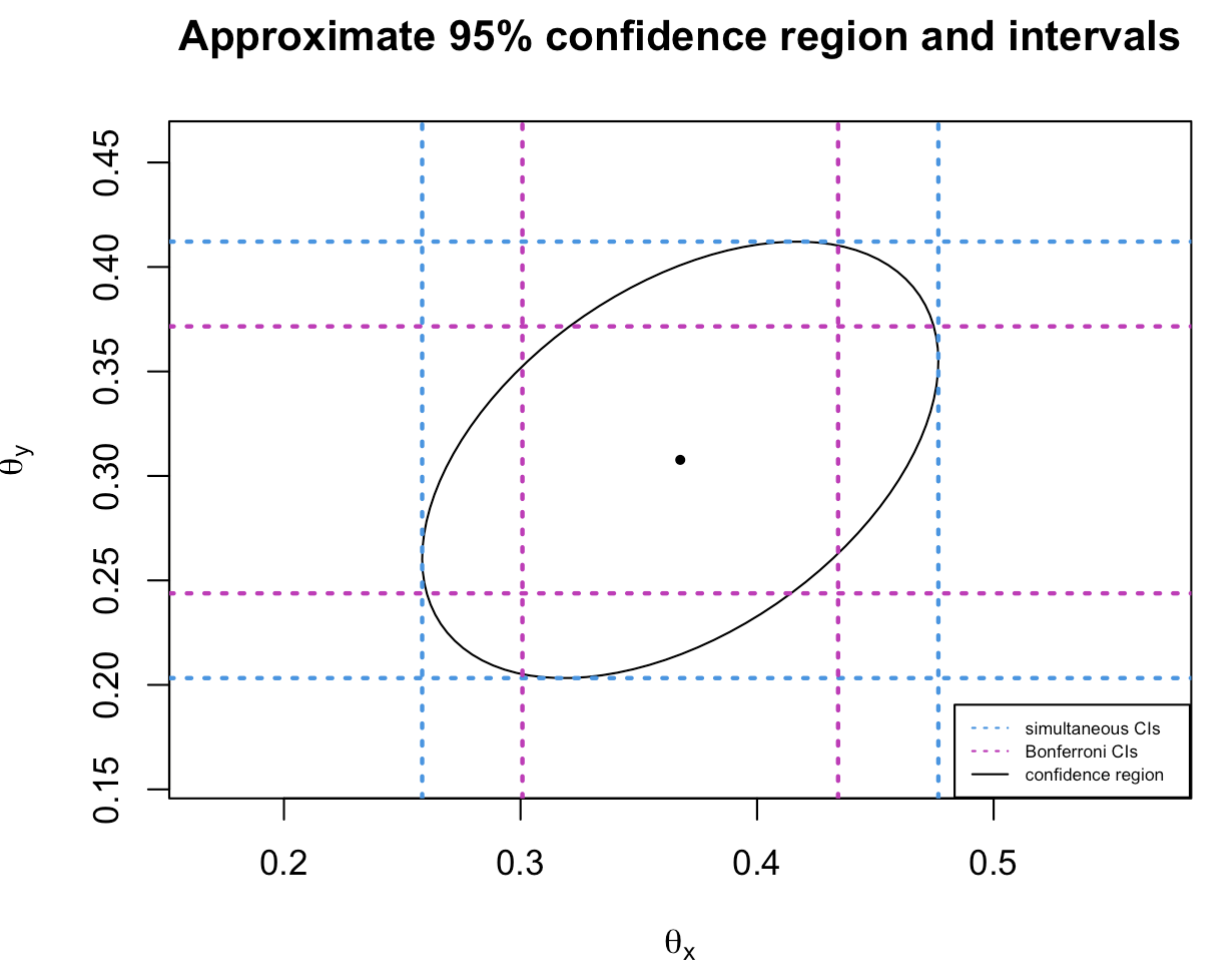}
 \caption{example 1}
 \end{subfigure}
 \hfill
 \begin{subfigure}[b]{0.48\textwidth}
\centering
\includegraphics[width=\textwidth]{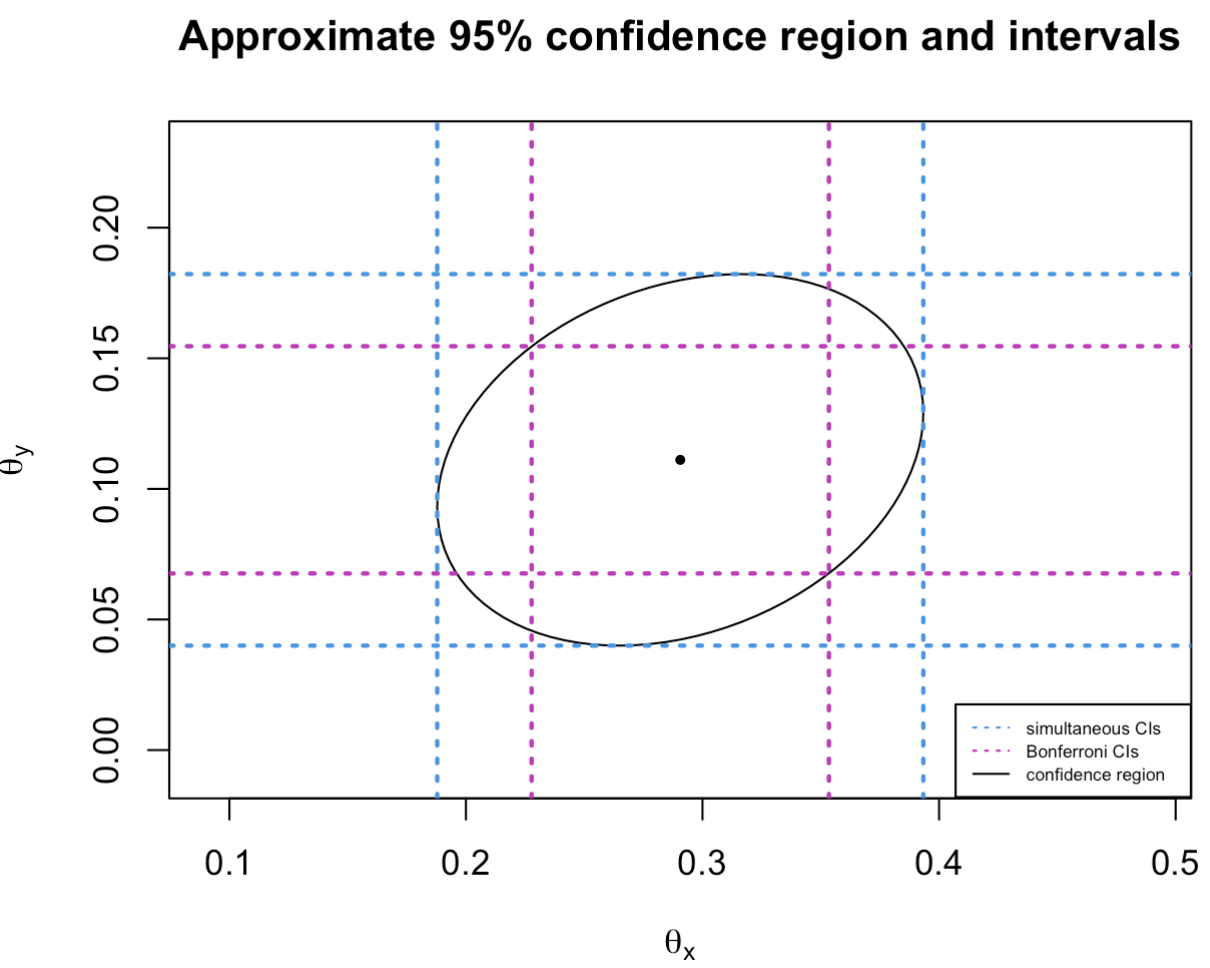}
\caption{example 2}
 \end{subfigure}
 \caption{The plots of the $95\%$ simultaneous confidence intervals of example $1$ and example $2$.}
 \label{f5}
 \end{figure}

\noindent \textbf {Example 1:}

For this example, we focus attention on the side effect {\it{muscle pain}}, $X$, and the side effect {\it{fever}}, $Y$. The results of these classifications are provided as a $2 \times 2$ table in Table \ref{t5} (a) above. For the post-test joint inference of $(\theta_x, \theta_y)$, we have to estimate $\theta_x$, $\theta_y$ and $p_{11}$ in order to estimate $\bm{\Sigma}$ in \eqref{5.31}. According to \eqref{2.3} and \eqref{2.4}, we calculate $\hat{\theta}_x=0.3675$, $\hat{\theta}_y=0.3077$, and by direct calculation $\hat{p}_{11}=\frac{n_{11}}{M^*}=\frac{25}{117}=0.2137$ and the corresponding $\hat{\rho}=0.4521$. Further, the $95\%$ confidence region is the ellipse shown in Figure \ref{f5} (a), which is centered at $(0.3675,0.3077)$ and the half-lengths of the major and minor axes are $0.1288$ and $0.0790$. By Theorem \ref{thrm2} and Theorem \ref{thrm3}, we obtain the $95\%$ simultaneous confidence intervals are $\theta_x \in [0.2584, 0.4766]$, $\theta_y \in [0.2032, 0.4121]$ and the $95\%$ Bonferroni confidence intervals are $\theta_x \in [0.3008, 0.4343]$, $\theta_y \in [0.2438, 0.3716]$.

In the following, we assume two possible illustrative scenarios which could have yielded the outcomes reported in table \ref{t5} (a) .

$\bullet$\  \underbar{Scenario I} ({\it{upon rejection of $H_0$}}) We reconstruct the optimal $(\alpha=0.05,\beta=0.09)$ sequential test of \eqref{1.1}. To that end, we assume here that $\theta_x^0=\theta_y^0=0.1$, then the pair of the test hypotheses is:
$$
H_0: \theta_x \le 0.1 \text{ and } \theta_y \le 0.1\ \ \text{against} \ \ H_1: \theta_x > 0.1 \text{ or } \theta_y >0.1 .
$$
According to \eqref{2.7} and \eqref{2.8}, assuming the termination occurring at $X$ boundary, $M_{x}\equiv 117< N^*$, $M^*=\min\{M_{x},N^*\}=117$ with $k^*_x=42$ and $S_{M_{x}}^y=36$. With $\alpha_x=0.025$ and the probability of Type II~error, $\beta_x=0.09$, in particular, assuming $\theta_x^1=0.16$, we obtain the maximal sample size $N^*_x=324$. Then for the same setup of the side effect $Y$, we obtain $N_y^*=324$, $k^*_y=42$. With no doubt that $N^*_x=N^*_y$, we have $N^*=\min\{N^*_x,N^*_y\}=324$. Since the estimated $\hat{\rho}=0.4521$, we calculate (retrospectively) the probabilities of Type~I error and Type~II error in practice as
$$
\Pr\left(\text{Type~I~error}\right)=0.0561 \quad \text{and} \quad \Pr\left(\text{Type~II~error}\right)=0.0208.
$$
Therefore, for this optimal $(\alpha \approx 0.05,\beta \approx 0.09)$ sequential test, with total sample size $N^* \equiv 324$ and the critical value $k^*_x=k^*_y=42$, we would stop the study and reject the null hypothesis.

$\bullet$\  \underbar{Scenario II} ({\it{a non-rejection of $H_0$}}) We reconstruct the optimal $(\alpha=0.05,\beta=0.092)$ sequential test of \eqref{1.1}. But now we assume $\theta_x^0=\theta_y^0=0.4$, then the pair of the test hypotheses is:
$$
H_0: \theta_x \le 0.4 \text{ and } \theta_y \le 0.4\ \ \text{against} \ \ H_1: \theta_x > 0.4 \text{ or } \theta_y >0.4.
$$
According to \eqref{2.7} and \eqref{2.8}, assuming $M=\min\{M_x,M_y\}>N^*\equiv 117$, we have $M^*=\min\{M,N^*\}=117$ and $S_{N^*}^x=43$, $S_{N^*}^y=36$. Assuming that $X$ and $Y$ have the same setup, we have $N^*=N_x^*=N_y^* =117$. With $\alpha_x=0.025$, we calculate the critical value $k^*_x=57$. To achieve the probability of Type~II error, $\beta_x=0.092$, in particular, we assume $\theta_x^1=0.55$. Then for the same setup of side effect $Y$, (that is, $\theta_y^0=0.4, \theta_y^1=0.55, \alpha_y=0.025, \beta_y=0.092$), we obtain $N^*_y=117$, $k^*_y=57$. Since the estimated $\hat{\rho}=0.4521$, we calculate (retrospectively) the probabilities of Type~I error and Type~II error in practice as
$$
\Pr \left(\text{Type~I~error}\right)=0.0402 \quad \text{and} \quad \Pr\left(\text{Type~II~error}\right)=0.0302.
$$
Therefore, for this optimal $(\alpha \approx 0.05,\beta \approx 0.092)$ sequential test, with total sample size $N^* \equiv 117$ and the critical value $k^*_x=k^*_y=57$, we would not reject the null hypothesis that the probabilities of participants exhibiting muscle pain and fever are both less than or equal to $0.4$.

Moreover, in this case, we may estimate the relative risk between the two side effects by $\hat{\gamma}=\hat{\theta}_x / \hat{\theta}_y=1.1944$. Based on Theorem \ref{thrm2} and Theorem \ref{thrm4}, utilizing the above estimators, we obtain the $95\%$ confidence interval is $[0.8740, 1.5149]$. Since this confidence interval includes the value of $\gamma=1$, it indicates that, for $\alpha=0.05$, we would not reject the null hypothesis, $H_0: \gamma=1$, that the relative risk between muscle pain and fever is $1$ (i.e. that $\theta_x=\theta_y$).

\noindent \textbf{Example 2:}

For this example, we focus attention on the side effect {\it{headache}}, $X$, and the side effect {\it{dizziness}}, $Y$. The results of these classifications are summarized as a $2 \times 2$ table in Table \ref{t5} (b). For the post-test joint inference of $(\theta_x, \theta_y)$, we have to estimate $\theta_x$, $\theta_y$ and $p_{11}$ in order to estimate $\bm{\Sigma}$ in \eqref{5.31}. According to \eqref{2.3} and \eqref{2.4}, we calculate $\hat{\theta}_x=0.2906$, $\hat{\theta}_y=0.1111$, and by direct calculation $\hat{p}_{11}=\frac{n_{11}}{M^*}=\frac{8}{117}=0.0684$ and the corresponding $\hat{\rho}=0.2529$. Further, the $95\%$ confidence region is the ellipse shown in Figure \ref{f5} (b), which is centered at $(0.2906,0.1111)$ and the half-lengths of the major and minor axes are $0.1055$ and $0.0670$. By Theorem \ref{thrm2} and Theorem \ref{thrm3}, we obtain the $95\%$ simultaneous confidence intervals are $\theta_x \in [0.1879, 0.3933]$, $\theta_y \in [0.0400, 0.1822]$ and the $95\%$ Bonferroni confidence intervals are $\theta_x \in [0.2278, 0.3534]$, $\theta_y \in [0.0676, 0.1546]$.

Similarly, in the following, we assume two possible illustrative scenarios which could have yielded the outcomes reported in Table \ref{t5} (b).

$\bullet$\  \underbar{Scenario I} ({\it{upon rejection of $H_0$}}) We reconstruct the optimal $(\alpha=0.05, \beta=0.09)$ sequential test of \eqref{1.1}. To that end, we assume here $\theta_x^0=\theta_y^0=0.1$, then the pair of the test hypotheses is:
$$
H_0: \theta_x \le 0.1 \text{ and } \theta_y \le 0.1\ \ \text{against} \ \ H_1: \theta_x > 0.1 \text{ or } \theta_y >0.1 .
$$
According to \eqref{2.7} and \eqref{2.8}, assuming the termination occurring at $X$ boundary, $M_{x}\equiv 117< N^*$, $M^*=\min\{M_{x},N^*\}=117$ with $k^*_x=33$ and $S_{M_{x}}^y=13$. With $\alpha_x=0.025$ and the probability of Type II~error, $\beta_x=0.09$, in particular, assuming $\theta_x^1=0.17$, we obtain the maximal sample size $N^*_x=243$. Then for the same setup of the side effect $Y$, we obtain $N_y^*=243$, $k^*_y=33$. With no doubt that $N^*_x=N^*_y$, we have $N^*=\min\{N^*_x,N^*_y\}=243$. Since the estimated $\hat{\rho}=0.2529$, we calculate (retrospectively) the probabilities of Type~I error and Type~II error in practice as
$$
\Pr\left(\text{Type~I~error}\right)=0.0472 \quad \text{and} \quad \Pr\left(\text{Type~II~error}\right)=0.0167.
$$
Therefore, for this optimal $(\alpha \approx 0.05,\beta \approx 0.09)$ sequential test, with total sample size $N^* \equiv 243$ and the critical value $k^*_x=k^*_y=33$, we would stop the study and reject the null hypothesis.

$\bullet$\  \underbar{Scenario II} ({\it{a non-rejection of $H_0$}}) We reconstruct the optimal $(\alpha=0.05, \beta=0.11)$ sequential test of \eqref{1.1}. But now we assume $\theta_x^0=\theta_y^0=0.31$, then the pair of the test hypotheses is:
$$
H_0: \theta_x \le 0.31 \text{ and } \theta_y \le 0.31\ \ \text{against} \ \ H_1: \theta_x > 0.31 \text{ or } \theta_y >0.31.
$$
According to \eqref{2.7} and \eqref{2.8}, assuming $M=\min\{M_x,M_y\}>N^*\equiv 117$, we have $M^*=\min\{M,N^*\}=117$ and $S_{N^*}^x=34$, $S_{N^*}^y=13$. Assuming that $X$ and $Y$ have the same setup, we have $N^*=N_x^*=N_y^* =117$. With $\alpha_x=0.025$, we calculate the critical value $k^*_x=46$. To achieve the probability of Type~II error, $\beta_x=0.11$, in particular, we assume $\theta_x^1=0.45$. Then for the same setup of side effect $Y$, (that is, $\theta_y^0=0.31, \theta_y^1=0.45, \alpha_y=0.025, \beta_y=0.11$), we obtain $N_y^*=117$, $k^*_y=46$. Since the estimated $\hat{\rho}=0.2529$, we calculate (retrospectively) the probabilities of Type~I error and Type~II error in practice as
$$
\Pr\left(\text{Type~I~error}\right)=0.0394 \quad \text{and} \quad \Pr\left(\text{Type~II~error}\right)=0.0288.
$$
Therefore, for this optimal $(\alpha \approx 0.05,\beta \approx 0.11)$ sequential test, with total sample size $N^* \equiv 117$ and the critical value $k^*_x=k^*_y=46$, we would not reject the null hypothesis that the probabilities of participants exhibiting muscle pain and fever are both less than or equal to $0.31$.

Moreover, similarly, we may estimate the relative risk $\hat{\gamma}=\hat{\theta}_x / \hat{\theta}_y=2.6154$. Based on Theorem \ref{thrm2} and Theorem \ref{thrm4}, utilizing the above estimators, we obtain the $95\%$ confidence interval is $[1.2578, 3.9730]$. Since the confidence interval exceeds $\gamma=1$, it indicates that, for $\alpha=0.05$, we would reject the null hypothesis, $H_0: \gamma=1$, that the relative risk between headache and dizziness is greater than $1$. In this case, we would conclude that people are more likely to exhibit dizziness than headache after they received COVID-19 vaccine.

\section{Summary and Discussion}

In this paper, we develop an $(\alpha,\beta)$-optimal sequential testing procedure for the early detection of two potential side effects of certain treatments. This sequential testing procedure does not require the specification of the correlation $\rho$, (if any) between the two potential side effects, nor any assumptions concerning it. Our procedure ensures that the actual probabilities of Type~I and Type~II errors would not exceed some desired levels of $(\alpha,\beta)$ for all possible values of $\rho$.

Since there is no assumption on the value of the correlation, we utilize the ('negative' version of the) multinomial distribution to derive the exact expression of the ASN and the variance of the 'stopping time' $M^*$. However, some tight bounds on the ASN are shown to hold following some simpler calculations, once some general information on $\rho$ is available. For instance, if these two side effects are independent (so that $\rho=0$), we have a simplified version of the ASN available. Following the basic analysis of the properties of the stopping time, we focus on the post-detection estimators of the model's parameters $\theta_x$ and $\theta_y$. We derive the exact formulas for calculating the expectation of the post-test (post-detection) estimators and similarly outline the derivation needed for calculating the corresponding variance.

To offset the computing time needed for the exact calculations, especially for values of $(\theta_x,\theta_y)$ in the close neighborhood of $(\theta_x^0,\theta_y^0)$, the asymptotic properties of the final sample size (i.e., the stopping time) are important to analyze. We derive the joint (bivariate) asymptotic normality of $(S_{M_\delta}^{z},M_\delta)^\prime$, for $z=x,y$, which is the crucial result for our subsequent analyses. Based on this asymptotic distribution, we approximate the probability distribution of the stopping time at each possible value in its support; a distribution that we then utilize to calculate the ASN, and the expectation and the variance of the post-test (post-detection) estimators, etc.

Moreover, the large sample consistency and the joint asymptotic normality of the post-detection estimators enable us to also construct the asymptotic normality of the estimated relative risk, $\gamma$, of the two side effects. In Section \ref{s6}, we presented two examples (based on real-life data) involving 'non-detection' and 'detection' situations of the side effects. In both examples, we apply our sequential testing procedure and calculate the post-test estimators, the corresponding joint confidence intervals, and also the estimated relative risk and its confidence interval. These examples clearly illustrate that our procedure performs well in the two different scenarios we assumed (such assumptions can be defined by specialists). We note that the nominal probabilities of Type~I and Type~II errors in these examples are less than the desired $(\alpha,\beta)$, since the critical values used do not utilize the correlation between these two side effects. The nominal values of these error probabilities are conservative in any applied situation. We conclude these two examples with the construction of a significance test of hypothesis concerning the relative risk, $\gamma$, between the two side effects we are interested in.

We point out that in some situations (e.g. $\theta_x \ll \theta_y$), the probability of $\operatorname{P}_{\uw{\theta}}(M_x>M_y)$ mentioned in Remark $1$ is close to $1$. It indicates that once we stop our observation process and reject the null hypothesis, we are likely to stop at the boundary $k_y^*+1$ corresponding to the side effect $Y$. In such a case, the one-dimensional optimal sequential test proposed in \citet{wang2024early} is sufficient to detect the potentially significant side effect (specifically, side effect $Y$). Similar conclusion can be obtained for the case when $\operatorname{P}_{\uw{\theta}}(M_x<M_y) \to 1$. 

Also note that if there are more than two potential side effects to account for, one can separate these side effects into multiple pairs (decision should involve input from the specialists), then applying our methods into each pair to have further analysis results. To match the large sample size requirement, our proposed test method can be sustainably applied to post-marketing surveillance data. 

In summary, we have demonstrated that our proposed sequential testing procedure is particularly useful for an early detection of multiple side effects, especially in emergency situations as during the rapid deployment of the COVID-19 vaccination campaign. Furthermore, the properties (asymptotic and 'finite-sample') of the the post-test estimates of the unknown prevalence of the potential side effects are very useful for any subsequent analysis.

\section{Technical Derivations}

{\bf{A.1}}
For side effect $X$, to construct the fixed-sample UMP test of
$$
H_0: \theta_x \le \theta_x^0 \ \ \text{against} \ \ H_1: \theta_x > \theta_x^0,
$$
since the indicator for side effect $X$ is Bernoulli random variable, we have the following properties:
\be \label{8.33}
\tilde{\alpha}:=\operatorname{P}_{\theta_x^0}(S_{N_x} >k_x)\leq \alpha,
\ee
and
\be \label{8.34}
\tilde{\beta}(\theta_x^1)= \operatorname{P}_{\theta_x^1}(S_{N_x} \le k_x)\le \beta.
\ee
Given corresponding $(\alpha, \beta,\theta_x^0,\theta_x^1)$, $(\theta_x^1>\theta_x^0)$,
we may simultaneously 'solve' equations \eqref{8.33} and \eqref{8.34} for $N_x$ and $k_x$ to obtain the optimal 'sample size', $N^*_x\equiv N_x(\alpha, \beta, \theta_x^0, \theta_x^1)$ and a corresponding 'critical test value', $k^*_x$, by either an iterative procedure utilizing \eqref{8.33} and \eqref{8.34} and the Binomial $pmf$ or by the standard Normal approximations to the Binomial probabilities\footnote{See conditions in $(1)-(2)$ of \citet{schader1989two}.} are given by,  
$$
N^*_x=\left[ \left(\frac{z_{1-\alpha}\sqrt{\theta_x^0(1-\theta_x^0)}+z_{\beta}\sqrt{\theta_x^1(1-\theta_x^1)}}{\theta_x^1-\theta_x^0}\right)^2 \right] , 
$$
and,
$$
k^*_x=\left[ N^*_x(z_{1-\alpha}\sqrt{\frac{\theta_x^0(1-\theta_x^0)}{N^*_x}}+\theta_x^0)-\frac{1}{2} \right], 
$$
where $[x]$ is the nearest integer value to $x$ and $z_p:=\Phi^{-1}(p)$, $\forall \ p \in (0,1)$ where $\Phi$ denotes the standard Normal $cdf$.

Similarly, we can obtain the corresponding $N^*_y$ and $k^*_y$ for side effect $Y$ by constructing the marginal fixed-sample UMP test for given the values of $(\alpha,\beta, \theta_y^0,\theta_y^1)$ as well.

\begin{lemma} \label{lem2}
$\Pi_{\operatorname{T^*_{seq}}}(\uw{\theta})$, the power function of optimal bivariate sequential test, it follows immediately from \eqref{2.9} that, $\forall \ \uw{\theta} \in \Theta_R$, is monotonically increasing with respect to $\theta_x$ and $\theta_y$.
\end{lemma}
\begin{prof}
\begin{align} \label{8.35}
\Pi_{\operatorname{T^*_{seq}}}(\uw{\theta})&=\operatorname{P}_{\uw{\theta}}\left(\operatorname{T^*_{seq}} \text{ reject } H_0 \right)=\operatorname{P}_{\uw{\theta}}\left(M \leq N^*\right)=1-\operatorname{P}_{\uw{\theta}}\left(M > N^*\right) \notag\\
&=1-\operatorname{P}_{\uw{\theta}}\left(S_{N^*}^x \le k^*_x \text{ and }S_{N^*}^y \le k^*_y \right) \notag\\
&=1-\sum_{z=0}^{\wu{k}}\sum_{i=0}^{k^*_x-z}\sum_{j=0}^{k^*_y-z}\frac{N^*!}{z!i!j!(N^*-z-i-j)!}p_{11}^z p_{10}^i p_{01}^j p_{00}^{N^*-z-i-j} .
\end{align}
To explore the tendency of the power function, we need to discuss separately with respect to $\theta_x$ or $\theta_y$. Once we fixed $\theta_x=\theta_x^\prime$, we may express the power function denoted as $\Pi_{\operatorname{T^*_{seq}}}(\theta_x^\prime,\theta_y) \equiv \Pi^\prime(\theta_y)$. Note that in this case, the multinomial probabilities of $\Pi^\prime(\theta_y)$ in \eqref{8.35} would be reduced to a binomial probabilities, a case which has been discussed in \citet{wang2024early}. Accordingly, it has been established there (see Theorem \ref{thrm1}) that $\Pi^\prime(\theta_y)$ is monotonically increasing with respect to $\theta_y$. Similarly, when we fix the value of $\theta_y$, we can establish the monotonicity of the power function with respect to $\theta_x$ as well. Hence, we have the monotonically
increasing property of the power function $\Pi_{\operatorname{T^*_{seq}}}(\uw{\theta})$ with respect to its ordinates $\theta_x$ and $\theta_y$.
\end{prof}

\begin{rem}  \label{rem3}
A key ingredient in the calculation of \eqref{3.11} was the probability mass function of $M \equiv\min\{M_x,M_y\}$. By utilizing the 'negative' version of the multinomial distribution in \eqref{2.2}, with $p_{00}=1-\theta_x-\theta_y+p_{11}$, $p_{10}=\theta_x-p_{11}$, $p_{01}=\theta_y-p_{11}$, $p_{11}=Cov(X,Y)+\theta_x\theta_y$, we obtain by direct derivation that
\be \label{8.36}
\operatorname{P}_{\uw{\theta}}\left(M=m\right)\equiv \operatorname{P}_{\uw{\theta}}\left(\min\{M_x,M_y\}=m\right) 
=A_1+A_2+A_3+A_4+A_5,
\ee
where
$$
A_1=\sum_{i=1}^{\min\{k^*_x+1,k^*_y\}}\sum_{j=0}^{\min\{k^*_y-i,m-k^*_x-1\}}A^{(1)}_{i,j}, \quad \quad
A_2=\sum_{i=0}^{\wu{k}}\sum_{j=0}^{\min\{k^*_y-i,m-k^*_x-1\}}A^{(2)}_{i,j}, 
$$
$$
A_3=\sum_{i=1}^{\min\{k^*_x,k^*_y+1\}}\sum_{j=0}^{\min\{k^*_x-z,m-k^*_y-1\}}A^{(3)}_{i,j}, \quad \quad
A_4=\sum_{i=0}^{\wu{k}}\sum_{j=0}^{\min\{k^*_x-i,m-k^*_y-1\}}A^{(4)}_{i,j},
$$
$$
A_5=\sum_{i=1}^{\wu{k}+1}A^{(5)}_{i}\mathbbm{1}\left[m \ge k^*_x+k^*_y+2-i\right], 
$$
and
$$
A^{(1)}_{i,j}=\frac{(m-1)!}{(i-1)!(k^*_x+1-i)!j!(m-k^*_x-1-j)!}p_{11}^i p_{10}^{k^*_x+1-i}p_{01}^j p_{00}^{m-k^*_x-1-j},
$$
$$
A^{(2)}_{i,j}=\frac{(m-1)!}{i!(k^*_x-i)!j!(m-k^*_x-1-j)!}p_{11}^i p_{10}^{k^*_x+1-i}p_{01}^j p_{00}^{m-k^*_x-1-j},
$$
$$
A^{(3)}_{i,j}=\frac{(m-1)!}{(i-1)!j!(k^*_y+1-i)!(m-k^*_y-1-j)!}p_{11}^i p_{10}^{j}p_{01}^{k^*_y+1-i} p_{00}^{m-k^*_y-1-j},
$$
$$
A^{(4)}_{i,j}=\frac{(m-1)!}{i!j!(k^*_y-i)!(m-k^*_y-1-j)!}p_{11}^i p_{10}^{j}p_{01}^{k^*_y+1-i} p_{00}^{m-k^*_y-1-j},
$$
$$
A^{(5)}_{i}=\frac{(m-1)!}{(i-1)!(k^*_x+1-i)!(k^*_y+1-i)!(m-k^*_y-k^*_x-2+i)!}p_{11}^i p_{10}^{k^*_x+1-i}p_{01}^{k^*_y+1-i} p_{00}^{m-k^*_y-k^*_x-2+i}.
$$
Here, $A_1$ and $A_2$ indicate that the process terminates because of exhibiting too many cases of side effect $X$, where $A_1$ represents the last observation exhibited side effect $X$ only and $A_2$ represents the last observation exhibited both side effects $X$ and $Y$; $A_3$ and $A_4$ indicate that the process terminates because of exhibiting too many cases of side effect $Y$, where $A_3$ represents the last observation exhibited side effect $Y$ only and $A_4$ represents the last observation exhibited both side effects $X$ and $Y$; $A_5$ is the special case when the random walk on the lattice stopped at the corner $(k^*_x+1,k^*_y+1)$.
\end{rem}

\begin{lemma} \label{lem3}
Let $X$ be a random variable with nature numbers, then
if $X$ is curtailed by fixed size $n$,
\be \label{8.37}
E \left(X\mathbbm{1}[X \le n] \right)=\sum_{i=1}^{n}\operatorname{P}(X \ge i)-n \operatorname{P}(X \ge n+1).
\ee
\end{lemma}

\begin{prof}
If $X$ is curtailed by fixed size $n$:
\begin{align*} 
E \left(X\mathbbm{1}[X \le n] \right)=&\sum_{i=0}^{n} i \operatorname{P}(X=i)  \\
=&\operatorname{P}(X=1)+2 \cdot \operatorname{P}(X=2)+3\cdot \operatorname{P}(X=3)+\dots+n\cdot \operatorname{P}(X=n) \\
=&\operatorname{P}(X=1)\\
+&\operatorname{P}(X=2)+\operatorname{P}(X=2) \\
+&\operatorname{P}(X=3)+\operatorname{P}(X=3)+\operatorname{P}(X=3) \\
&\vdots \\
+&\operatorname{P}(X=n)+\operatorname{P}(X=n)+\operatorname{P}(X=n)+\dots+\operatorname{P}(X=n) \\
=&\sum_{i=1}^{n}\operatorname{P}(X \ge i)-n \operatorname{P}(X \ge n+1).
\end{align*}
\end{prof}

\begin{lemma} \label{lem4}
$\forall \ \uw{\theta}\in \Theta_R$, if the correlation between side effect $X$ and side effect $Y$, $\rho \ge 0$, the stopping time $M_x$ and the stopping time $M_y$ are positively associated; if $\rho \le 0$, $M_x$ and $M_y$ are negatively associated; if $X$ and $Y$ are independent, $M_x$ and $M_y$ are independent.
\end{lemma}

\begin{prof}
Notice that $Cov(X,Y)=\rho\sqrt{\theta_x(1-\theta_x)\theta_y(1-\theta_y)}$, if $\rho \ge 0$, $Cov(X,Y) \ge 0$; if $\rho \le 0$, $Cov(X,Y) \le 0$. Since ${(X_i, Y_i)}_{i=1}^{n}$ are i.i.d. two-dimensional random variables, we have $\forall \ n_1,\ n_2 \in \mathbbm{N^+}$,
$$
Cov\left(S_{n_1}^x, S_{n_2}^y\right)=Cov\left(\sum_{i=1}^{n_1}X_i, \sum_{j=1}^{n_2} Y_j\right)=\sum_{i=1}^{\min\{n_1,n_2\}}Cov\left(X_i,Y_i\right).
$$
Hence, if $\rho \ge 0$, we have $Cov\left(S_{n_1}^x, S_{n_2}^y\right)\ge 0$; if $\rho \le 0$, we have $Cov\left(S_{n_1}^x, S_{n_2}^y\right) \le 0$. Since by definitions of $M_x$ and $M_y$ in \eqref{2.6} can be represent as
$$
M_x=\sum_{n=0}^{\infty}\mathbbm{1}\left[S_n^x \le k^*_x\right] \quad \text{and} \quad M_y=\sum_{n=0}^{\infty}\mathbbm{1}\left[S_n^y \le k^*_y\right],
$$ 
where $\mathbbm{1}[S_n^x \le k^*_x]$ and $\mathbbm{1}[S_n^y \le k^*_y]$ are nonincreasing functions of $S_n^x$ and $S_n^y$.

Hence since '$-M_x$' and '$-M_y$' are nondecreasing functions of $S_n^x$ and $S_n^y$ and $Cov(-M_x,-M_y)=Cov(M_x,M_y)$, we have if $\rho \ge 0$, $M_x$ and $M_y$ are positively associated; if $\rho \le 0$, $M_x$ and $M_y$ are negatively associated (for properties of associated random variables see \citet{esary1967association} and \citet{joag1983negative}). On the other hand, if $X$ and $Y$ are independent, since $M_x$ and $M_y$ are Borel-measurable functions of $X$ and $Y$, we have $M_x$ and $M_y$ are independent.
\end{prof}

\begin{lemma} \label{lem5}
Let $X$ be a random variable with nature numbers, then
$$
E\left(X^2\right)=2\sum_{i=1}^{\infty}\left(i-\frac{1}{2}\right)\operatorname{P} \left(X \ge i \right).
$$
If $X$ is curtailed by fixed size $n$, then
$$
E \left(X^2\mathbbm{1} \left[X \le n \right] \right)=2\sum_{i=1}^{n}\left(i-\frac{1}{2}\right)\operatorname{P}\left(X \ge i \right)-n^2 \operatorname{P}\left(X \ge n+1 \right).
$$
\end{lemma}

\begin{prof}
For	$X$ is a random variable with nature numbers, since
\begin{align*}
\frac{E \left(X^2\right)+E \left(X\right)}{2}=&E \left( \frac{X(X+1)}{2} \right) 
=\sum_{i=1}^{\infty}\frac{i(i+1)}{2}\operatorname{P}\left(X=i\right) \\
=&\sum_{i=1}^{\infty} \left(\sum_{j=1}^{i} j \right)\operatorname{P} \left(X=i\right) 
=\sum_{j=1}^{\infty} j \sum_{i=j}^{\infty} \operatorname{P}\left(X=i\right) \\
=&\sum_{j=1}^{\infty} j  \operatorname{P} \left(X \ge j\right).
\end{align*}

If $X$ is curtailed by fixed size $n$, since
\begin{align*}
E \left( \frac{X(X+1)}{2}\mathbbm{1} \left[X \le n \right] \right)=&\sum_{j=1}^{n} j  \operatorname{P} \left(X \ge j \right)\\
=&\operatorname{P} \left(X=1\right) \\
+&\operatorname{P} \left(X=2\right)+2\operatorname{P} \left(X=2\right) \\
+&\operatorname{P} \left(X=3\right)+2\operatorname{P}\left(X=3\right)+3\operatorname{P}\left(X=3\right) \\
&\vdots \\
+&\operatorname{P}\left(X=n\right)+2\operatorname{P}\left(X=n\right)+3\operatorname{P}\left(X=n\right)+\dots+n\operatorname{P}\left(X=n\right) \\
=& \sum_{i=1}^{n} i \operatorname{P} \left(X \ge i \right)-\sum_{i=1}^{n}  i \operatorname{P} \left(X \ge n+1 \right) \\
=& \sum_{i=1}^{n} i \operatorname{P}\left(X \ge i \right)-\frac{n(n+1)}{2} \operatorname{P} \left(X \ge n+1 \right),
\end{align*}
and by \eqref{8.37} of Lemma \ref{lem3}, we have
\begin{align*}
E \left(X^2\mathbbm{1} \left[X \le n \right] \right)=&2E \left( \frac{X(X+1)}{2}\mathbbm{1} \left[X \le n \right] \right)-E \left(X\mathbbm{1} \left[X \le n \right] \right) \\
=&2 \left[\sum_{i=1}^{n} i \operatorname{P}\left(X \ge i \right)-\frac{n(n+1)}{2} \operatorname{P}\left(X \ge n+1 \right) \right]\\
-& \left[\sum_{i=1}^{n}\operatorname{P}\left(X \ge i \right) - n \operatorname{P} \left(X \ge n+1 \right)\right] \\
=&2\sum_{i=1}^{n}\left(i-\frac{1}{2}\right)\operatorname{P}\left(X \ge i \right)-n^2\operatorname{P}\left(X \ge n+1 \right).
\end{align*}
\end{prof}

{\bf{A.2}}
Note that in the evaluation of $\operatorname{P}_{\uw{\theta}}\left( M_\delta=m \right)$ in \eqref{8.36} of Remark \ref{rem3} above, involve the evaluation of 
$$
\operatorname{P}_{\uw{\theta}}\left( M_\delta=m \right)= A_1+A_2+A_3+A_4+A_5 $$
can be written as, by Lemma \ref{lem1}, as $\delta \to 0$,
\begin{align} \label{8.38}
&\operatorname{P}_{\uw{\theta}}\left( M_\delta=m \right) \equiv\operatorname{P}_{\uw{\theta}}\left(\min\{M_x^\delta,M_y^\delta\}=m \right) \notag\\
=&\int_{-\infty}^{k^\delta_y+0.5}\int_{m-0.5}^{m+0.5} \phi_2\left( u, w \mid \bm{\mu_2}, \bm{V_2} \right) dw du
+\int_{-\infty}^{k^\delta_x+0.5}\int_{m-0.5}^{m+0.5} \phi_2\left( u, w \mid \bm{\mu_3}, \bm{V_3} \right) dw du+A_5.
\end{align}
Note that the first term in \eqref{8.38} indicates the probability when the last observation exhibiting side effect $X$ causes the sum of observations that exhibited side effect $X$ to attain the critical value $k^\delta_x+1$ but $S_{M_\delta}^y<k^\delta_y+1$ (corresponding to $A_1$ and $A_2$ in \eqref{8.36}). Also note the second term in \eqref{8.38} indicates the probability when the last observation exhibiting side effect $Y$ causes the sum of observations that exhibited side effect $Y$ to attain the critical value $k^\delta_y+1$ but $S_{M_\delta}^x<k^\delta_x+1$ (corresponding to $A_3$ and $A_4$ in \eqref{8.36}).

However, whenever $\delta$ is small (as $\delta \to 0$), the value of $A_5$ in \eqref{8.38} which presents the probability of the last terminal observation exhibiting both side effect $X$ and side effect $Y$ causes both critical values $k^\delta_x+1$ and $k^\delta_y+1$ to be attained, which can be approximated as 
$$
A_5 =p_{11} \cdot\int_{k^\delta_x-0.5}^{k^\delta_x+0.5}\int_{k^\delta_y-0.5}^{k^\delta_y+0.5} \phi_2\left( u, w \mid \bm{\mu_4}, \bm{V_4}\right) dw du,
$$
where
$$
\bm{\mu_4}=\begin{pmatrix}[1.5]
	\left(m-1 \right)\theta_x \\
	\left(m-1 \right)\theta_y
\end{pmatrix}, \  \bm{V_4}=\begin{pmatrix}[1.5]
	\left(m-1 \right)\theta_x(1-\theta_x) & \left(m-1 \right)\rho\sqrt{\theta_x(1-\theta_x)\theta_y(1-\theta_y) } \\
	\left(m-1 \right)\rho\sqrt{\theta_x(1-\theta_x)\theta_y(1-\theta_y) } & \left(m-1 \right)\theta_y(1-\theta_y). 
\end{pmatrix}
$$
where $\phi_2( u, w \mid \bm{\mu_4}, \bm{V_4})$ denotes the bivariate normal distribution of $(S_{m-1}^x, S_{m-1}^y)^\prime$.

When we transform $(S_{m-1}^x, S_{m-1}^y)^\prime$ as the bivariate normal with mean $(0,0)^\prime$ and variance-covariance matrix as $\begin{pmatrix}1 &\rho \\ \rho &1\end{pmatrix}$, we obtain the difference values between upper bounds and lower bounds of the integration of $S_{m-1}^x$ and $S_{m-1}^y$ are, as $\delta \to 0$,
$$
\frac{k_x^\delta+0.5-\left(m-1 \right)\theta_x}{\sqrt{\left(m-1 \right) \theta_x \left(1-\theta_x \right)}}-\frac{k_x^\delta-0.5-\left(m-1 \right)\theta_x}{\sqrt{\left(m-1 \right) \theta_x \left(1-\theta_x \right)}}=\frac{1}{\sqrt{\left(m-1 \right) \theta_x \left(1-\theta_x \right)}} \to 0,
$$
since $m\geq \wu{k}_\delta +1 \to\infty$ and similarly,
$$
\frac{k_y^\delta+0.5-\left(m-1 \right)\theta_y}{\sqrt{\left(m-1 \right) \theta_y \left(1-\theta_y \right)}}-\frac{k_y^\delta-0.5-\left(m-1 \right)\theta_y}{\sqrt{\left(m-1 \right) \theta_y \left(1-\theta_y \right)}}=\frac{1}{\sqrt{\left(m-1 \right) \theta_y \left(1-\theta_y \right)}} \to 0.
$$
It indicates that the value of $A_5$ is negligible as $\delta \to 0$. Therefore, as $\delta \to 0$, we may use the approximated expression of $\operatorname{P}_{\uw{\theta}}(M_\delta=m)$, that is
\begin{align*} 
&\operatorname{P}_{\uw{\theta}}\left( M_\delta=m \right) \equiv\operatorname{P}_{\uw{\theta}}\left(\min\{M_x^\delta,M_y^\delta\}=m \right) \notag\\
=&\int_{-\infty}^{k^\delta_y+0.5}\int_{m-0.5}^{m+0.5} \phi_2\left( u, w \mid \bm{\mu_2}, \bm{V_2} \right) dw du
+\int_{-\infty}^{k^\delta_x+0.5}\int_{m-0.5}^{m+0.5} \phi_2\left( u, w \mid \bm{\mu_3}, \bm{V_3} \right) dw du,
\end{align*}
to simplify the calculations.

\printbibliography

\end{document}